\documentclass[12pt]{article}
\usepackage{amsmath,amssymb,bm,epsf,epsfig,graphicx,dsfont,caption,subcaption}
\usepackage[section]{placeins}
\usepackage{hyperref}
\input epsf.sty
\numberwithin{equation}{section}

\def\ket#1{|#1 \rangle}

\def\l{\left}

\def \be {\begin{eqnarray}}
\def \ee {\end{eqnarray}}
\def \bdm {\begin{displaymath}}
\def \edm {\end{displaymath}}

\def\del {\partial}
\def\0{\nonumber}

\def\l{{\rm \lambda}}

\def\lS{\lambda_{\rm S}}
\def\lB{\lambda_{\rm B}}
\newcommand{\cn}{\mathop{\rm cn}\nolimits}

\def \rar {\rightarrow}
\def \la {\langle}
\def \ra {\rangle}

\def \sn {\text{sn}}
\def \cn {\text{cn}}
\def \dn {\text{dn}}

\begin{document}
\begin{center}
 {\Large \bf $\,$\\
\vskip2cm
BCFT moduli space in level truncation}
\vskip.5cm

\vskip 0.4cm

{  Mat\v{e}j Kudrna\footnote{Email:
kudrnam at fzu.cz}$^{(a)}$, Carlo Maccaferri\footnote{Email:
maccafer at gmail.com}$^{(b)}$}  
\vskip 1 cm

$^{(a)}${\it {Institute of Physics of the ASCR, v.v.i.} \\
{Na Slovance 2, 182 21 Prague 8, Czech Republic}}
\vspace{.5cm}

$^{(b)}${\it Dipartimento di Fisica, Universit\'a di Torino and INFN, Sezione di Torino\\
Via Pietro Giuria 1, I-10125 Torino, Italy}\\

\end{center}

\vspace*{6.0ex}

\centerline{\bf Abstract}
\bigskip
We propose a new non-perturbative method to search for marginal deformations in level truncated open string field theory.
Instead of studying the flatness of the effective potential for the marginal field (which is not expected to give a one-to-one
parametrization of the BCFT moduli space),
we identify a new non-universal branch of the tachyon potential which, from known analytic examples, is expected to parametrize the marginal flow
in a much larger region of the BCFT moduli space.
By a level 18 computation in Siegel gauge we find an increasingly flat effective potential in the non-universal sector,
connected to the perturbative vacuum and we confirm that the coefficient of the marginal field ($\lambda_{\rm SFT}$)  has a maximum compatible with
the value where the solutions stop existing in the standard Sen-Zwiebach approach. At the maximal reachable level the effective potential still deviates from flatness
for large values of the tachyon, but the Ellwood invariants stay close to the correct BCFT values on the whole branch and the full periodic moduli space of the cosine deformation is covered.
\newpage

\baselineskip=15pt

\tableofcontents

\section{Introduction}
Since the first analytic solution for the tachyon vacuum has been found \cite{Schnabl}, our understanding of the classical non-perturbative aspects of open string field theory (OSFT) has importantly progressed  (see \cite{FK-rev, martin-rev, yuji-rev} for reviews) and  consistent conformal boundary conditions on the worldsheet (D-branes) have been cast into exact OSFT solutions  \cite{EM}. However we still  lack of a clear string-field-theoretic understanding
of how a generic OSFT solution in target space codify new, possibly unknown, conformal  boundary conditions on the worldsheet.

In absence of a constructive analytic understanding of the space of solutions of OSFT,  level truncation (LT) is  an important predictive tool which allows to numerically scan
for the D-branes landscape, starting from a chosen D-brane system \cite{MSZ, Michi, Ising}. In order for LT to work in practice one has to fix a gauge and the most common choice is
Siegel gauge. This is also  very natural  since perturbation theory in Siegel gauge  gives a direct geometric decomposition of the moduli space of Riemann surfaces  with boundaries \cite{cover}.
However it is essentially unknown whether every known OSFT classical solution can be gauge-transformed to Siegel gauge.

In this paper we non-trivially test Siegel gauge level truncation against large marginal deformations of the initial D-brane configuration. Concrete examples of these `far' solutions in OSFT
have been presented in \cite{simple-marg} and in \cite{EM} building on \cite{TT, simple, KOS, EM-gauge, IKT}. It is therefore an important question  whether  we can find these solutions in Siegel gauge as well.

It is well known that a Siegel gauge solution for marginal deformations can be written as a perturbative expansion in the coefficient
of the exactly marginal state $c j(0)\ket0\equiv cj$ as
\be
\Psi(\lS)=\lS\, cj-\lS^2\,\frac{b_0}{L_0}(cj*cj)+\lS^3\,\frac {b_0}{L_0}\left\{\frac{b_0}{L_0}(cj*cj)\, ,\,cj\right\}_*+\cdots\;.\label{exact-siegel}
\ee
For practical purposes this expression is however somehow formal because of the growing complexity of the involved world-sheet geometries arising from the combined use of the three-strings vertex and the Siegel gauge propagator \cite{giddings}. More importantly,  there is no obvious reason to expect that a power series in
$\lS$ has a large enough radius of convergence in the Fock space  to cover the BCFT moduli space associated with the marginal direction $j$.
A non-perturbative approach to this problem was initiated by Sen and Zwiebach \cite{large} and further
explored in \cite{senT, kurs, TT-branch, large2}\footnote{A similar problem, from a complementary different perspective, has been addressed, up to level (3,9), in \cite{KL1}.}. A major common point of these works is that, by parametrizing the marginal solutions with the VEV of the marginal field $\lambda_{\rm SFT}\equiv \lS$, no solution can be found after a certain critical value of the BCFT parameter $\lambda_{\rm BCFT}\equiv\lB$ which, in the case of the cosine deformation of a free boson
at the self-dual radius, is close
to the point where the initial Neumann boundary condition becomes Dirichlet, \cite{large2}\footnote{By taking into account also negative values for $\lS$,
this approximatively covers half of the full periodic moduli
space of the cosine deformation, or equivalently two fundamental domains out of four.}.

Recently it has been shown in \cite{large-moduli}
that in the case of the physically equivalent analytic solution \cite{simple-marg}, which is directly expressed in terms of the BCFT modulus $\lB$,
a power series in $\lS$ would necessarily stop converging
at finite radius, simply because $\lS$ is  not an injective function of $\lB$ and therefore the dependence on $\lS$ is multi-valued. In particular if we express $\lS$ as a function of $\lB$ we find
that it starts growing up to a maximal value and then decreases to zero at large $\lB$, see figure 2 of \cite{large-moduli}.

Precisely the same
behavior was in fact observed long time ago in \cite{toy} by analyzing
a very similar problem in $\phi^3$ scalar field theory expanded around its exact lump solution. Since the lump  has an obvious translational
modulus one can interpret (minus) the difference
between the lump and its translation as an exact solution of the shifted action. It was then noticed that the VEV acquired by the translational
Goldstone mode (the SFT parameter $\lS$)    is not
an injective function of the amount of translation (the BCFT modulus $\lB$), but it increases up to a maximum and then starts decreasing and  relaxes to zero at infinite
translation, in the case of a decompactified transverse direction. At the same time the coefficient of the tachyon
mode (responsible for the instability of the lump) was also tracked down as a function of the separation and shown to be injectively related
to $\lB$:
the tachyon mode starts growing quadratically in $\lB$ and then asymptotes monotonically to a constant value which, when the transverse direction is non-compact, is just the tachyon
coefficient of the stable vacuum of the $\phi^3$ theory expanded around the lump (the tachyon vacuum), see figure 5 of \cite{toy}. One therefore expects to find a flat direction
in the effective tachyon potential  which is in one-to-one correspondence with the full lump moduli space.

Going back to OSFT, the tachyon coefficient of the analytic solution \cite{simple-marg} was also shown to asymptote (from above, after a local maximum), to a constant positive value
 with a very similar qualitative behavior as the $\phi^3$ toy model \cite{toy}, see figure 3 of \cite{large-moduli}.
In both examples the VEV of
the tachyon is a much better coordinate in the solution moduli space than the VEV of the marginal field.
This motivated us to explore marginal deformations in Siegel gauge by searching for flat directions in the (non-universal) tachyon effective potential.
We summarize our results by giving the plan of the paper.

In section 2 we test our idea in Zwiebach's $\phi^3$ toy model compactified on a circle. We find an exact lump solution and we expand the action around it. The circle
compactification guarantees a discrete spectrum for the eigenfuctions of the kinetic operator around the lump and the whole program of level truncation becomes concretely addressable.  We first
study the effective potential for the marginal field and find that, at low levels, there is a fairly flat branch connected to the perturbative vacuum which however ends and meets with another
branch which is connected to the the tachyon vacuum. This is perfectly parallel to the situation in string field theory. However, by just adding the first two degenerate massive levels (which would
be part of the continuous spectrum in the decompactification limit), two more branches are generated,
one of the two being fairly flat. Already at level 4 this new marginal branch gets superimposed with the original marginal branch. This is precisely what we expect since the marginal coefficient of
the known exact solution starts growing, reaches a maximum (roughly corresponding to the end of the first branch) and then decreases (the new branch).  Interestingly, we observe that the first marginal branch, connected to the trivial vacuum, is on-shell only up to the point where it meets with the new marginal branch appearing at level 2:  the remaining part of this  branch doesn't belong to the moduli space of the lump.
We then analyze the tachyon effective potential of the same toy model and  find that already at low level it has a single flat direction ending at a critical tachyon VEV which limits to  the tachyon vacuum in the decompactification limit.
This is again as expected since the tachyon
coefficient of the exact solution of the toy model, in the relevant region of moduli space,  is an injective function of the lump translation.

In section 3 we study the analogous problem in Siegel gauge OSFT on the concrete example of the cosine deformation at the self-dual radius, which is $SU(2)$-dual to the translation of a $D0$-brane. Already at level 2 we find a non universal branch
in the tachyon potential which is connected to the perturbative vacuum and which is reasonably flat for (very) small tachyon VEV $t$. We look at the value of the marginal parameter $\lS$ on this
newly found tachyon branch and, at level 5, we find that it starts showing a maximum as a function of $t$. By improving the solutions up to $L=18$ we confirm that $\lS$ has indeed a maximum.
We  numerically relate
the tachyon VEV $t$ with the BCFT modulus $\lB$, by  fitting the Ellwood invariants \cite{Ellwood} against their expected value from BCFT, \cite{large2, KMS}. We find that the  Ellwood invariants are remarkably close to the Ishisbashi states coefficients of the known BCFT boundary state, even when (at the reachable level) the equation of motion for the tachyon is quite far from being satisfied. For completeness we repeat our analysis in the standard marginal approach, to check the consistency between the two approaches in the common region of moduli space. Comparing the two approaches we find  that the last part of the marginal branch in the marginal approach is off-shell, precisely as it happens in the $\phi^3$ toy-model.

We end with some comments and future directions.

\section{Toy model}  \label{sec:toy}
\begin{figure}
\centering
\includegraphics[width=8cm]{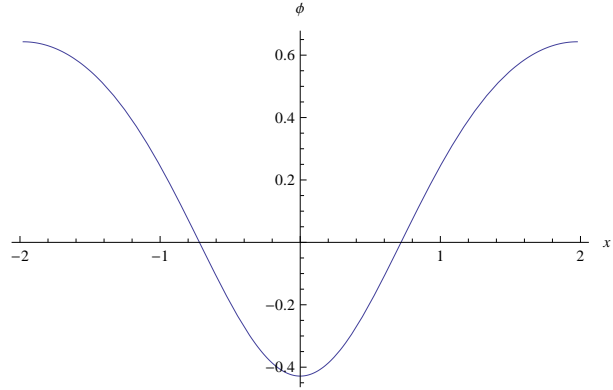}
\caption{Profile of the lump for $m=\frac{3}{4}$.}\label{fig:lump_profile}
\end{figure}

In this section we briefly explain our strategy in the simple $\phi^3$ toy model described in \cite{toy}, but compactified on a circle of radius $R$.
We start with a one-dimensional scalar with action given by
\begin{equation} \label{toy action}
S=\int_{-\pi R}^{\pi R} dx \left[ -\frac{1}{2}\left( \frac{\del \phi}{\del x}\right)^2 -V(\phi) \right].
\end{equation}
The cubic potential is chosen as in \cite{toy}
\be
V(\phi)=\frac13(\phi-1)^2\left(\phi+\frac12\right).\label{V-toy}
\ee
Then we have  to search for solutions to the field equation
\begin{equation}\label{toy phi equation}
\frac{d^2 \bar \phi}{dx^2}-V'(\bar \phi)=0,
\end{equation}
exibiting the prescribed periodicity
\be
\bar\phi(x+2\pi R)=\bar\phi(x).
\ee
To find the lump solution at given radius $R$ we follow the mechanical analog in the inverted tachyon potential and we fix the maximal value of the lump profile $\phi_{\rm max}$.
In the corresponding mechanical model, this is the
point where the kinetic energy is vanishing. The mechanical system will  oscillate around the perturbative vacuum $\phi=0$ in the inverted tachyon potential and the period of
oscillation is to be identified with the compactification circumference in the field theory. Therefore fixing the compactification radius implicitly fixes one integration constant in the field equation.
The other integration constant will correspond to translations (the lump modulus). Having implicitly defined $\phi_{\rm max}$, it is convenient to express the (shifted) potential in terms of its
roots.
\begin{equation}
V(\phi)-V(\phi_{\rm max})=\frac{1}{3}\left(\phi-a_1\right)\left(\phi-a_2\right)\left(\phi-a_3\right),
\end{equation}
where the constant part $V(\phi_{\rm max})$ doesn't enter in the field equations. By simple match, the roots $a_i$'s satisfy
\begin{eqnarray}
a_1+a_2+a_3&=&\frac{3}{2}, \nonumber \\
a_1 a_2+a_1 a_3+a_2 a_3&=&0. \label{toy roots}
\end{eqnarray}
The searched-for periodic solution
is  given (up to translations) in terms of the sn Jacobi elliptic function
\begin{equation}
\bar\phi(x)=a_1+(a_2-a_1)\ \sn^2\left(\frac{\sqrt{a_3-a_1}}{\sqrt{6}}x\,{\Big|}\,\frac{a_2-a_1}{a_3-a_1}\right).
\end{equation}
It is convenient to define the elliptic modulus
\be
m&\equiv&\frac{a_2-a_1}{a_3-a_1},\\
m&\in& [0,1],\nonumber
\ee
and, solving (\ref{toy roots}), to express the roots as
\begin{eqnarray}
a_1 &=& \frac{-1-m+\Delta_m}{2\Delta_m},\\
a_2 &=& \frac{-1+2m+\Delta_m}{2\Delta_m},\\
a_3 &=& \frac{2-m+\Delta_m}{2\Delta_m},
\end{eqnarray}
where $\Delta_m \equiv \sqrt{m^2-m+1}$. The exact periodic solution is therefore given by
\begin{equation}
\bar\phi(x)=\frac{-1-m+\Delta_m+3m\ \sn^2 \left(\frac{1}{2\sqrt{\Delta_m}}x\,{\Big|}\,m\right)}{2\Delta_m},
\end{equation}
see figure \ref{fig:lump_profile}.
The compactification radius $R$ is directly related to the real periodicity of sn$^2$ by
\be
R(m)=\frac{2\sqrt{\Delta_m}K(m)}{\pi},
\ee
where $K(m)$ is the complete elliptic integral of the first kind. Notice that for $m=0$ we find $R(0)=1$ and for $m=1$ we have $R(1)=\infty$. The lump doesn't form for $R\leq1$.
By translational invariance in the $x$ direction  we get the expected one-dimensional moduli space of the lump
\be \label{lump-moduli}
\bar\phi_{\lB}(x)=\bar\phi(x-\lB).
\ee

\begin{figure}
\centering
\includegraphics[width=8cm]{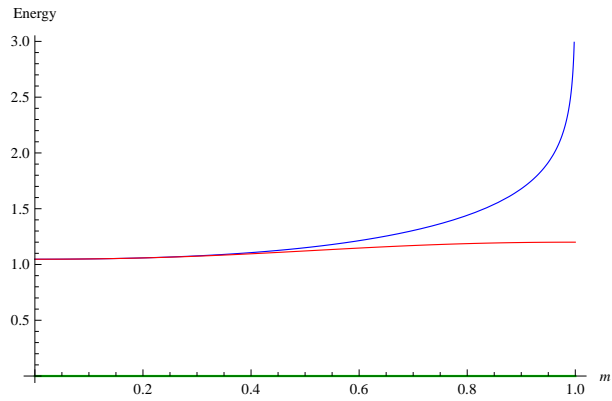}
\caption{The energy of $\bar \phi$ lump (red), $\phi=0$ solution (blue) and $\phi=1$ solution (green) as a function of $m$.}\label{fig:toy_energy}
\end{figure}

Together with the lump, analogous to a $D0$-brane, there are also the  solutions $\phi=\phi_{\rm pv}=0$ (perturbative vacuum, analogous to the $D1$-brane) and $\phi=\phi_{\rm tv}=1$
(analogous to the tachyon vacuum). Their energies, computed from the action (\ref{toy action}) are plotted in figure \ref{fig:toy_energy} as a function of $m\in[0,1]$. Notice that
as $m\to0$ or equivalently $R\to1$ the lump and the perturbative vacuum energies asymptote each other, as in string theory. However, differently from string theory, there is really no lump at $R=1$
\be
\bar\phi_{m=0}=0=\phi_{\rm pv}.
\ee

\subsection{Fluctuations around the lump}

Now we expand the action (\ref{toy action}) around the lump $$\phi(x)=\bar \phi(x)+\psi(x),$$ and we span the fluctuations with the eigenfunctions of the induced kinetic operator
\be
\psi(x)=\sum_n \xi_n\psi_n(x),
\ee
\begin{equation}
-\frac{d^2\psi_n}{dx^2}+V''\left(\bar\phi(x)\right)\psi_n(x)=M^2_n\psi_n(x).
\end{equation}
This turns the shifted action into a cubic function of the coefficients $\xi_n$, which is analogous (up to the absence of the  winding modes and most notably the descendants of primaries)   to the level truncated OSFT action on a D0-brane on a circle. Truncating the action to a maximal $n$ gives a system with a finite number of degrees of freedom and the solutions to the field equations can be found numerically by solving a system of $2n+1$ coupled quadratic  equations.
The Schr\"{o}dinger problem for finding the fluctuations and their mass reads, in our case
\begin{eqnarray}\label{toy Schrodinger}
-\frac{d^2\psi_n}{dx^2}+\frac{-1-m+3m\ \sn^2 \left(\frac{1}{2\sqrt{\Delta_m}}x{\Big|}m\right)}{\Delta_m}\psi_n(x)&=&M^2_n\psi_n(x),\\
\psi_n(x+2\pi R(m))&=&\psi_n(x).\nonumber
\end{eqnarray}
By making the change of variables $y=x/(2\sqrt\Delta_m)$ ($\psi_n(y)\equiv \psi_n(x(y))$) it can be written as a Lam\'e equation for $j=3$ with periodic boundary conditions
\begin{eqnarray}
-\frac{d^2\psi_n}{dy^2}+j(j+1)\,m\ \sn^2 (y|m)\psi_n(y)&=&{\cal E}_n\psi_n(y),\quad (j=3)\\
\psi_n(y+2K(m))&=&\psi_n(y).\nonumber
\end{eqnarray}
The eigenvalues ${\cal E}_n$ have to be determined by imposing the prescribed periodicity and it is related to the mass $M_n$ of the fluctuation as
\be
{\cal E}_n=4(1+m+\Delta_m M^2_n).
\ee

This Schr\"{o}dinger problem admits solutions in terms of periodic Lam\'e functions (also known as ellipsoidal harmonics, see e.g.  \cite{BatemanIII}, 15.5). In appendix \ref{app:Lame} we present some solutions (in particular the exact eigensystem for the bound states). It turns out however that our Mathematica code is more efficient with the following straightforward numerical treatment.
We choose the standard  basis of functions on the circle $\left\{\cos \frac{kx}{R},\,\sin \frac{kx}{R}\right\}$ and
we expand the $\sn^2$ function in this basis. Since the kinetic operator is even with respect to the reflection $x\rar -x$,  its eigenfuctions are either even or odd
\be
\psi_n^{(e)}(x)&=&\sum_{k\geq0}\psi^{(e)}_{nk} \cos \frac{kx}{R},\0\\
\psi_n^{(o)}(x)&=&\sum_{k\geq1}\psi^{(o)}_{nk} \sin \frac{kx}{R}.\label{decomp}
\ee
By putting a cutoff on the maximal harmonic number $k$, the kinetic operator in (\ref{toy Schrodinger}) is converted into a finite dimensional matrix (block-diagonal in the even/even and odd/odd
subspaces)
and by computing its eigenvalues (denoted as $M_{(e/o),n}^2$)  and eigenvectors (the coefficients $\psi_{nk}^{(e/o)}$)  we get approximate solutions of (\ref{toy Schrodinger}),
which quickly stabilize (for fixed eigenvalue) as we increase the cutoff in the
harmonics.

The mass squared of the fluctuations up to $n=10$ for $m=\frac{3}{4}$ ($R\sim1.30$) is shown in table \ref{tab:mass squared}.
We find that the first state is even and tachyonic, the second state is odd and massless and the third is even and massive.
Then we  find pairs of even and odd states with degenerate positive mass. In the decompactification limit $m\to1$ the first
three states will form a discrete spectrum while the other pair-degenerate ones
will form a continuos spectrum, as described in \cite{toy}.

\begin{table}\nonumber
\centering
\begin{tabular}{|l|ll|l|}\hline
$n$& $M_{(e),n}^2$  & $M_{(o),n}^2$ &$\frac{n^2}{R^2}$ \\\hline
0  & -1.22375 & -       & 0       \\
1  & 0.53037  & 0       & 0.58860 \\ \hline
2  & 1.95509  & 1.95509 & 2.35441 \\
3  & 4.85145  & 4.85145 & 5.29743 \\
4  & 8.95575  & 8.95575 & 9.41766 \\
5  & 14.2458  & 14.2458 & 14.7151 \\
6  & 20.7165  & 20.7165 & 21.1897 \\
7  & 28.3659  & 28.3659 & 28.8416 \\
8  & 37.1934  & 37.1934 & 37.6706 \\
9  & 47.1986  & 47.1986 & 47.6769 \\
10 & 58.3813  & 58.3813 & 58.8604 \\\hline
\end{tabular}
\caption{Mass squared of the first solutions to the Schr\"{o}dinger equation for $m=\frac{3}{4}$, $R\sim 1.30$. The first three eigenvalues correspond to the bound states of the Schr\"{o}dinger potential. The rest of the spectrum corresponds to the non-bound states and will become a continuum in the decompactification limit. The shown digits are stable under addition of higher harmonics in our numerical determination of the spectrum. For comparison we show the mass spectrum of the plane waves on the right, to which our spectrum asymptotes for large eigenvalues.}
\label{tab:mass squared}
\end{table}

Now we can determine the action of the fields living on the lump from (\ref{toy action}) by taking terms proportional to the perturbations expanded into even and odd normalized eigenfunctions
\be
\psi(x)=\sum_{n\geq0}\left(\xi_n^{(e)}\psi_n^{(e)}(x)+\xi_n^{(o)}\psi_n^{(o)}(x)\right),
\ee
\begin{eqnarray}
S[\xi]&=&\int_{-\pi R}^{\pi R} dx\left[\frac{1}{2}\psi\left(-\frac{d^2}{dx^2}+V''(\bar \phi)\right)\psi+\frac{1}{3}\psi^3 \right]\\
&=&\sum_{n\geq0}\frac12\left(\xi_n^{(e)}M_{(e),n}^2\xi_n^{(e)}+\xi_n^{(o)}M_{(o),n}^2\xi_n^{(o)}\right)\nonumber\\
&+&\sum_{n,m,p\geq0}\left(\frac13C_{nmp}^{(e,e,e)}\xi_n^{(e)}\xi_m^{(e)}\xi_p^{(e)}+C_{nmp}^{(e,o,o)}\xi_n^{(e)}\xi_m^{(o)}\xi_p^{(o)}\right)\nonumber,
\end{eqnarray}
where the interaction constants are given by cubic integrals of the normalized eigenfuctions
\be
C_{nmp}^{(e,e,e)}&=&\int_{-\pi R}^{\pi R}dx \,\psi_n^{(e)}(x)\psi_m^{(e)}(x)\psi_p^{(e)}(x),\\
C_{nmp}^{(e,o,o)}&=&\int_{-\pi R}^{\pi R}dx \,\psi_n^{(e)}(x)\psi_m^{(o)}(x)\psi_p^{(o)}(x),\nonumber
\ee
and can be numerically computed to any desired precision using (\ref{decomp}).

This action is our starting point for ``level truncation'' studies in the  $(L,3L)$ scheme, where by ``level'' $L$ we  mean keeping all states with $n\leq L$. Notice that in this field theory
toy model the number of states at level $L$ is given by $2L+1$. From now on, as a concrete representative example, we show results for $m=\frac{3}{4}$, corresponding to
$R\left(\frac{3}{4}\right)=1.30343$.

\begin{figure} \label{fig:toy tach and lambda}
 \begin{minipage}[b]{0.5\linewidth}
 \centering
 \resizebox{3.2in}{1.7in}{\includegraphics[scale=1]{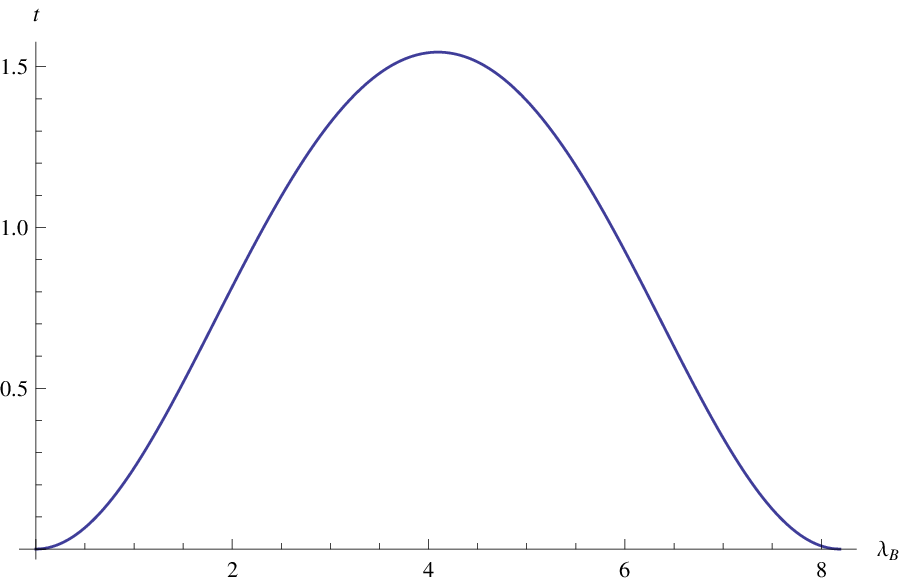}}
 \end{minipage}
 \hspace{0.5cm}
 \begin{minipage}[b]{0.5\linewidth}
 \centering
 \resizebox{3.2in}{1.7in}{\includegraphics[scale=1]{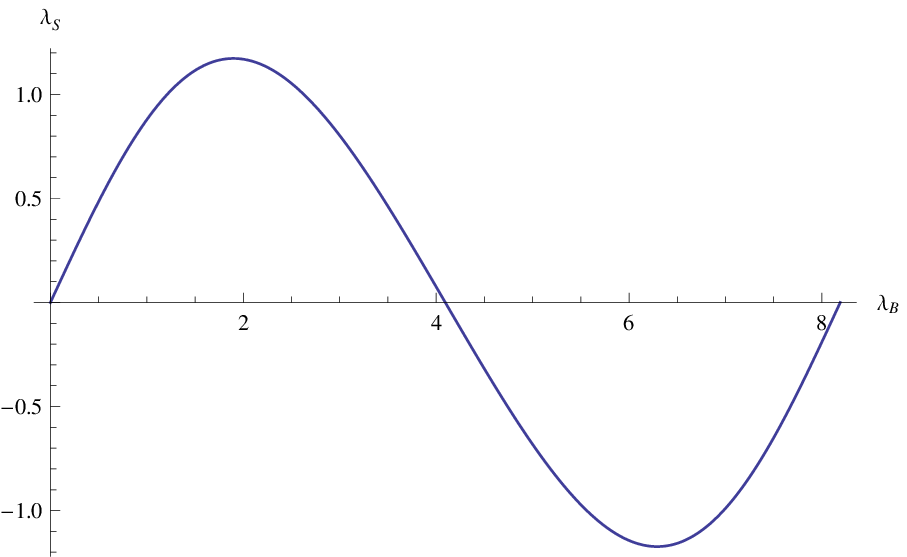}}
 \end{minipage}
\caption{{\small On the left: Plot of the tachyon coefficient $t(\l_B)$ as a function of the physical displacement $\lB$ (\ref{t}). On the right: Plot of the marginal coefficient $\l_S(\l_B)$ (\ref{l}). }}\label{fig:toy-coeff}
\end{figure}

In the following our aim will be to test how, in this toy model, level truncation  reconstructs the exact ``marginal'' solution
\be
\Psi^{\rm marg}_{\lB}(x)=\bar\phi(x-\lB)-\bar\phi(x),\label{psi_marg}
\ee
describing a finite translation of the lump, from the perspective of a lump centered at the origin. In figure \ref{fig:toy-coeff} we plot the  tachyon and marginal coefficients of the above
marginal solution, as function of the physical amount of translation $\lB$ (the ``BCFT'' parameter), which can be calculated by contracting the solution (\ref{psi_marg}) with the corresponding normalized eigenfunctions
\be
t(\l_B)&=&\int_{-\pi R}^{\pi R} dx\,\Psi^{\rm marg}_{\lB} (x) \,\psi_0^{(e)}(x)\label{t}\\
\l_S(\l_B)&=&\int_{-\pi R}^{\pi R} dx\,\Psi^{\rm marg}_{\lB} (x)\, \psi_1^{(o)}(x),\label{l}
\ee
where  $\psi_{0/1}^{(e/o)}$ are respectively the (normalized) tachyon and  Goldstone mode of the background lump.

It should be noticed that the behavior for large $\lB>\pi R$ is sensibly different from the available OSFT example of \cite{large-moduli} (see fig. 2 and 3 there). On the other hand the plot for $\lB<\pi R$ is in qualitative agreement with OSFT. There is a  simple explanation for this: OSFT is a gauge theory and the solution has to come back to itself only up to a (large) gauge transformation, upon  winding around the circular moduli space. In particular the OSFT solution describing the translation of a D0-brane does not depend  on the compactification radius and the circular moduli space is formed because the boundary condition changing operators translating by $2\pi R$ are in fact genuine fields of the theory (they are the open string winding modes) from which it is possible to build the above-mentioned gauge transformations, \cite{EM, simple-marg}. The toy model, on the other hand, does not have any gauge symmetry and the solution has to be strictly periodic in $\lB$.

\subsection{Marginal approach in the toy model} \label{sec:toy marginal}

\begin{figure}
 \begin{minipage}[b]{0.5\linewidth}
 \centering
 \resizebox{3.2in}{1.7in}{\includegraphics[scale=1]{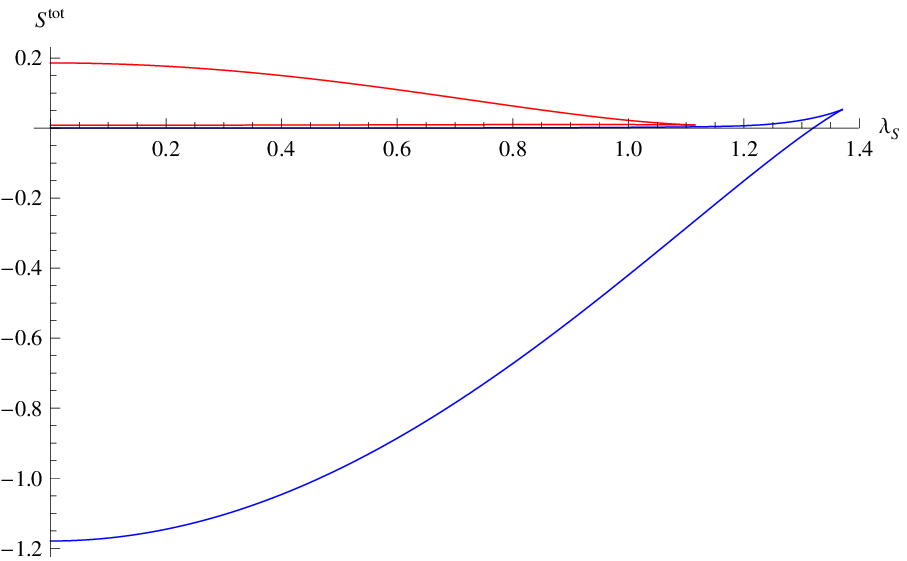}}
 \end{minipage}
 \hspace{0.5cm}
 \begin{minipage}[b]{0.5\linewidth}
 \centering
 \resizebox{3.2in}{1.7in}{\includegraphics[scale=1]{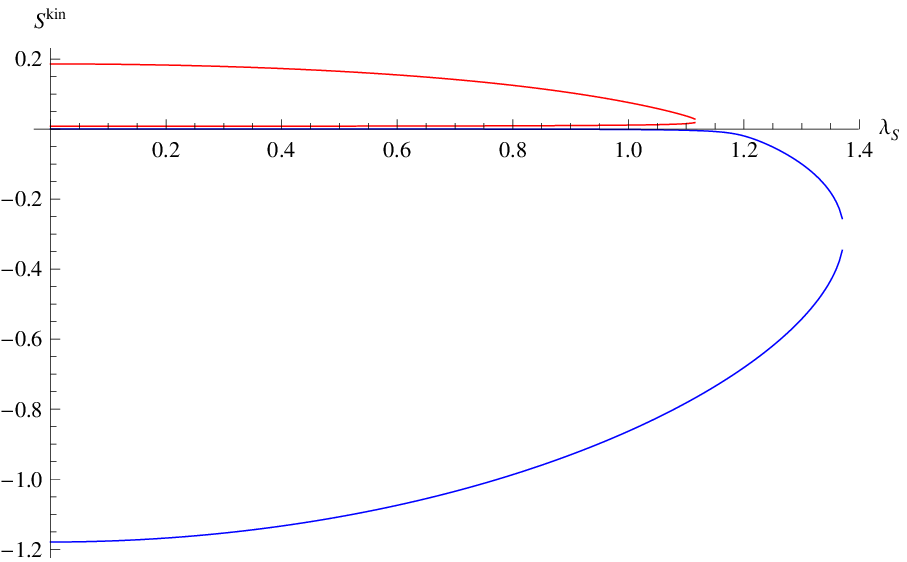}}
 \end{minipage}
 \caption{{\small On the left: Action of the four branches of solutions in marginal approach at $L=2$, computed from the full action.
 On the right: Action from kinetic term at $L=2$.}}\label{fig:toy mar energy}
\end{figure}

In the ``marginal'' approach we fix the coefficient $\l_S$ of the massless field and we leave the corresponding equation unsolved.
At level 1 (meaning that we keep the tachyon, the marginal field and the first massive state) we find only two branches of real solutions, depicted in blue in figure \ref{fig:toy mar energy}.
This is qualitatively the same branch structure that we observe in OSFT: the solutions end at a finite value of the VEV for the marginal field and the lump moduli space is not fully covered.
However already at level 2  (that is, including the first two pair-degenerate massive fluctuations, which would be part of the continuous spectrum in the decompactification limit) we find two other branches of solutions, depicted in red in figure \ref{fig:toy mar energy}.
In total two of the branches are truly marginal.  The other two non-flat branches do not satisfy the missing equation of motion and they are not vacua of the theory. The longer off shell branch is connected to the $\phi=1$ solution (tachyon vacuum) and the shorter one to the $\phi=0$ solution (D1-brane).
By increasing the level, this branch structure remains and, already at level 3, the two marginal branches get essentially superimposed with respect to the overall scale. At level 10 the action of the marginal branches is vanishing within  15 digit precision.

\begin{figure}
\centering
\includegraphics[width=10cm]{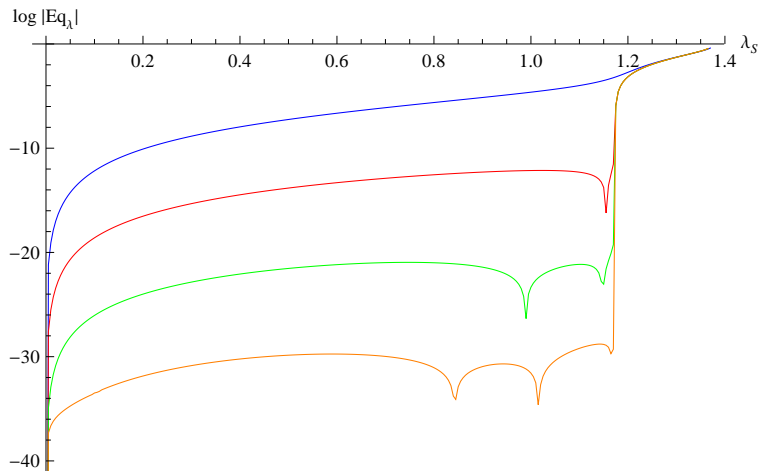}
\caption{Absolute value of the missing equation of the first marginal branch in logarithmic scale. Blue line corresponds to $L=2$, red to $L=4$, green to $L=6$ and orange to $L=8$.}\label{fig:toy lambda equation}
\end{figure}

\begin{figure}
\centering
\includegraphics[width=8cm]{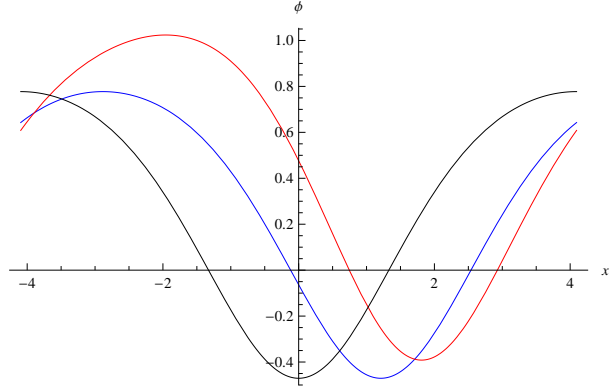}
\caption{Profile of the lump (black), reconstructed shifted lump $\bar \phi(x)+\psi(x)$ for $\lS=1$ (blue) and false shifted lump for $\lS=1.35$ (red) at level 10.}\label{fig:toy profile}
\end{figure}

This branch structure is interesting and reveals a new unexpected feature.
The long branches end approximately at $\lS=1.37$, the short branches at $\lS=1.17$. Therefore the two marginal branches meet before the longer branch ends. The critical value at which they meet corresponds with high precision (already at level 2) to the maximal VEV of the marginal field for the exact solution (\ref{psi_marg}).
Surprisingly  the remaining part of the long marginal branch is off-shell! When we look at the missing equation for the marginal field we indeed find that it is satisfied very well up to $\lS=1.17$ and then it quickly grows by several orders  (see figure \ref{fig:toy lambda equation}). This is further confirmed by reconstructing the shifted lump $\bar\phi(x-\l_B)$ from the numerical solution and noticing that after the critical value $\l_S\sim 1.17$  the  `solution'  sensibly deviates from a shifted lump, see figure \ref{fig:toy profile}. The data strongly suggest that for the exact solution
at $L\to \infty$ all four branches meet at the same point and the off-shell part of the long marginal branch  smoothly joins the off-shell short branch, although they are distinct at finite level.

\begin{figure}
\centering
\includegraphics[width=10cm]{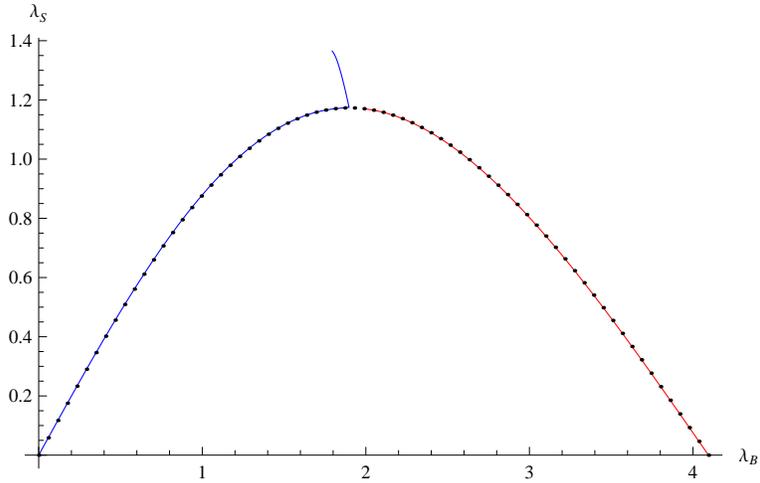}
\caption{Relation between the marginal field $\lS$ and position of the lump $\l_B$ in the marginal approach at level 10. The blue line is the first marginal branch (including its off-shell part, the small blue ``hair") and the red line is the second branch. The two branches cover half of the circle, the other half is covered by solutions with negative $\lS$. The black dots show the exact relation given by (\ref{l}), which perfectly agrees with the level truncation data.}\label{fig:toy mar position}
\end{figure}
We can also easily obtain the physical parameter $\l_B$ (amount of translation) as a function of the marginal parameter $\l_S$. To do this we find the minimum in the reconstructed translated lump profile
in the two marginal branches, see figure \ref{fig:toy mar position}.

\begin{figure}
\centering
\begin{subfigure}[t]{0.47\textwidth}
\includegraphics[width=\textwidth]{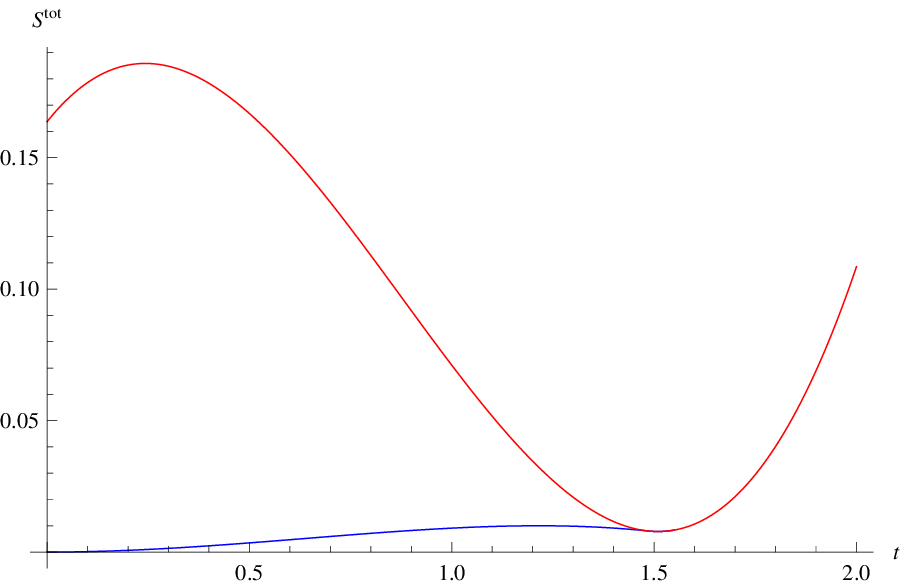}
\end{subfigure}\qquad
\begin{subfigure}[t]{0.47\textwidth}
\includegraphics[width=\textwidth]{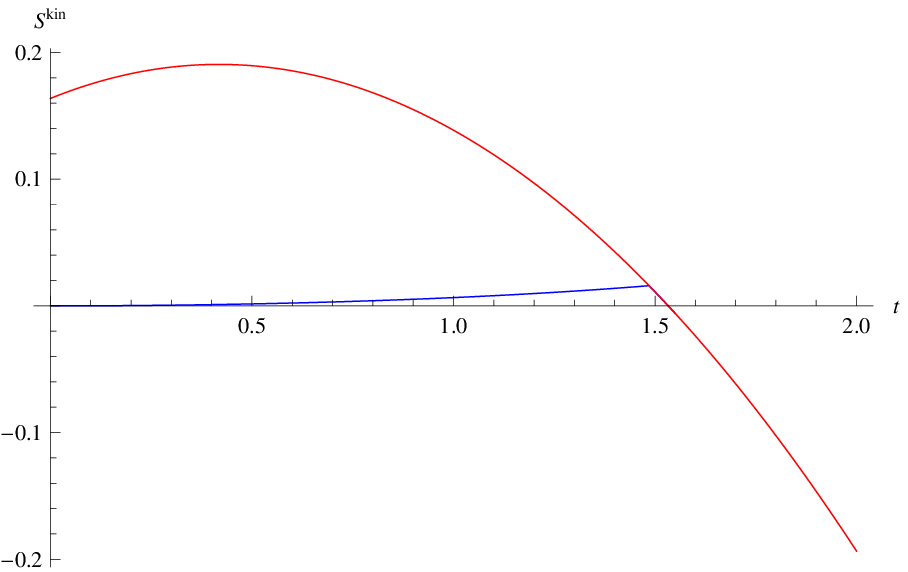}
\end{subfigure}
\caption{Full action (left) and action from kinetic term (right) in the tachyon approach at level 2. At higher levels the marginal branch would be undistinguishable from the $t$-axis.}\label{fig:toy tach energy}
\end{figure}
\subsection{Tachyon approach in the toy model} \label{sec:toy tachyon}

In the ``tachyon'' approach we fix the value of the tachyon and we leave its field equation unsolved.

\begin{figure}
\centering
\includegraphics[width=8cm]{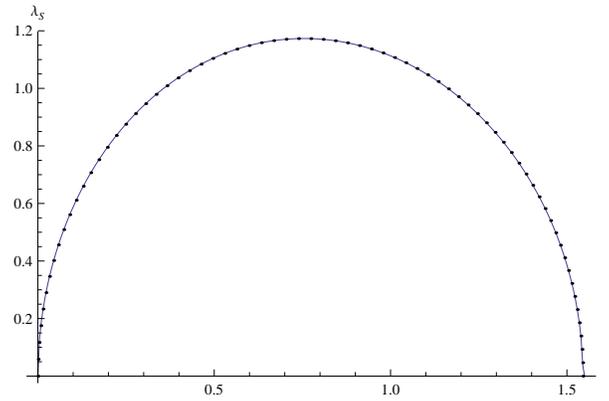}
\caption{The marginal field $\lS$ as a function of tachyon $t$ for the marginal branch. The blue line shows the numeric data at level 10 and the black dots the exact relation.}\label{fig:toy tach lambda}
\end{figure}

\begin{figure}
\centering
\includegraphics[width=8cm]{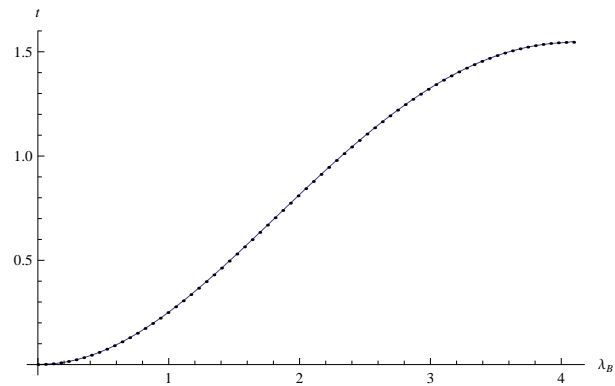}
\caption{Position of the lump $\lB$ in the tachyon approach at level 10.
The black dots show the exact relation given by (\ref{t}).}\label{fig:toy tach position}
\end{figure}

In this approach we find four branches of real solutions. There are two degenerate branches that have the same energy and differ by the sign of the marginal field (depicted in blue in figure \ref{fig:toy tach energy}). There is one branch with positive energy (depicted in red in figure \ref{fig:toy tach energy}) and one branch with negative energy that connects perturbative and tachyon vacuum (not shown in the figure). The two marginal branches end and meet with the unphysical branch. This branch structure is  simpler than in the marginal approach and the single flat direction we find accounts for the whole lump moduli space (including the other specular branch where $\l_S$ has opposite sign). 
In figure \ref{fig:toy tach lambda} we see that the non-injective dependence of $\lS$ on $t$ is captured in a single branch and in figure \ref{fig:toy tach position} we check that the relation between the lump position $\lB$ and the tachyon coefficient $t$ allows to cover the full moduli space up to $\lB=\pi R$. The $\lB>\pi R$ part of moduli space is covered by the mirror branch with $\lS<0$.

Therefore in the tachyon approach  a much larger region of the lump moduli space (in fact the whole moduli space) is captured by a single branch of the tachyon effective potential. This is clearly a much better situation for level truncation studies and this is the approach we are now going to test for Siegel-gauge level-truncated OSFT.

\section{Siegel Gauge OSFT}

In this section we study the tachyon approach to marginal deformations in OSFT. We take the relevant BCFT to be a free boson at the self-dual radius ($R=1$) with Neumann boundary conditions and no Wilson line. We choose the $\cos X$ marginal deformation to allow comparison with the most recent results from \cite{large2}.

Given a numerical solution $\Psi$, its  boundary state can be computed using the generalized Ellwood invariants discussed in \cite{KMS}.
The primary operators of a compact free boson at $R=1$, which define the Ishibashi states, can be classified by the $SU(2)$ symmetry, see for example \cite{Recknagel}. An holomorphic operator with $SU(2)$ spin $(j,m)$ has conformal weight $h=j^2$ and left-handed momentum $k=m$.
Leaving the details of the construction to \cite{KMS}, the coefficients of the first Virasoro Ishibashi states are explicitly computed from a given solution $\Psi$ (together with a tachyon vacuum solution $\Psi_{\rm tv}$) as
\begin{eqnarray}
E_n&=& 2\pi i \la E[c\bar c \cos(n X) V^{(1-n^2/4)}]|\Psi-\Psi_{\rm tv}\ra, \nonumber \\
W_n&=& 2\pi i \la E[c\bar c \cos(n \tilde X) V^{(1-n^2/4)}]|\Psi-\Psi_{\rm tv}\ra, \nonumber \\
D_1&=& 4\pi i \la E[c\bar c \del X  \bar \del X]|\Psi-\Psi_{\rm tv}\ra, \nonumber \\
H  &=&-4\pi   \la E[c\bar c \del X  \sin \bar X]|\Psi -\Psi_{\rm tv}\ra, \label{Elw definition}
\end{eqnarray}
where $V^{(h)}=e^{2i\sqrt{1-h}Y}$ is the analytic continuation of a bulk plane wave in the $Y$-BCFT with Dirichlet boundary conditions, such that
$\langle e^{2i\sqrt{1-h}Y}(0)\rangle^{\rm Dir}_{\rm disk}=1$. In principle higher weights invariants can be also considered, but they are usually not very well behaved in level truncation.
The above defined invariants should match the coefficients of the Ishibashi states which are given (for example) in  \cite{Recknagel} for a generic $SU(2)$ deformation. For the cosine deformation we have
\begin{eqnarray}
E_0 &=& 1, \nonumber \\
E_1 &=& -\sin \pi\lB, \nonumber \\
E_2 &=& \sin^2\pi\lB, \nonumber \\
W_1 &=& \cos\pi\lB, \nonumber \\
W_2 &=& \cos^2\pi\lB,\nonumber \\
D_1 &=& \cos 2\pi\lB, \nonumber \\
H   &=& -\frac{1}{\sqrt{2}}\sin 2\pi\lB.\label{Elw expected}
\end{eqnarray}
From these invariants we see that the moduli space is periodic with the identification $\lB \sim \lB+ 2$. $\l_B=0$ is the initial Neumann boundary condition with no Wilson line. At $\l_B=1/2$ we find Dirichlet boundary conditions $X=\pi$, then at $\l_B=1$ we have Neumann boundary conditions with a constant Wilson line $\omega=\pi$ and at $\l_B=3/2$ we have Dirichlet boundary conditions $X=0$.

The low level fields excited by the Siegel gauge solution are the tachyon and the marginal field
\be
\Psi=t\,c_1\ket0+\l_S\,c_1 j_{-1}\ket0+\textrm{(higher levels)}.
\ee
In the tachyon approach we solve all fields (including the marginal fields) in terms of the tachyon VEV $t$, while in the marginal approach we solve in term of the marginal VEV $\lS$.
In the numerical calculations we use  $(L,3L)$ level truncation scheme. The string field is spanned by Virasoro generators in the universal matter sector, by $\alpha$-oscillators acting on momentum primaries in the free boson sector and by twisted Virasoros in the $SU(1,1)$ basis of the ghost sector \cite{ZwiebachSU11,GaiottoRastelli}. Additionally we take  the string field to be even with respect to  twist-symmetry and $X$-parity. 
We leave the detailed numerical algorithms to other references (see \cite{RastelliZwiebach} for conservation laws for cubic vertices, appendix of \cite{GaiottoRastelli} for the evaluation of the action and for solving the equations of motion, \cite{KMS} for Ellwood invariants.) In \cite{KS} some of the algorithms we used  will  be described in detail.

\begin{figure}[h]
\centering
\includegraphics{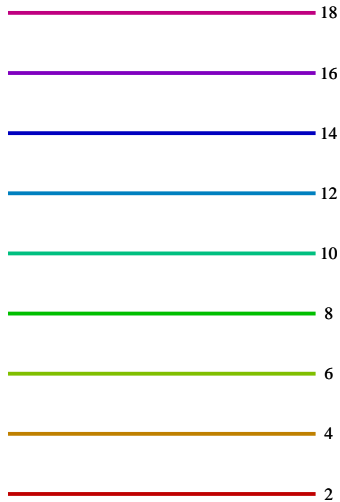}
\caption{Colors of levels in the figures.}\label{fig:Colors}
\end{figure}

    With a C++ code running on parallel clusters we can reach level 18, where we have 34842 fields. We computed the solution both at even and odd levels, however we present only data at even levels to avoid overcrowding\footnote{As in \cite{large2} there is no qualitative difference between data at odd and even levels}.  The colors of the curves in the figures correspond to the level and they follow the light spectrum from red at level 2 to purple at level 18 (see figure \ref{fig:Colors}).

Finally let us comment on our extrapolations to infinite level. We assume that all the quantities behave asymptotically as a series in $1/L$ so we fit them with a function 
\begin{equation}
a_0+\sum_{k=1}^N\frac{a_k}{L^k}.
\end{equation}
The order $N$ is usually 1 or 2 because higher orders typically give unstable results. Fully estimating the errors of the $L\to\infty$ extrapolations is not really possible. We easily  compute a ``statistical" error  by the standard deviation of different extrapolations with varying parameters (typically the number of included levels, as in \cite{KMS}). However there are important  unknown ``systematic'' errors that come from the fact the some quantities can have different asymptotics at high levels or, as it often happens with Ellwood invariants, anomalous behaviour coming from Pad\'{e}-Borel approximation. Unsurprisingly we find that the errors grow with $t$ and $\lS$, respectively in the tachyon and marginal approaches.

For quantities that follow a stable pattern (for example the energy, $\lS$, $E_0$) the statistical error is usually quite small (of order 1\% or less). These quantities can be dominated by the systematic errors, for example the potential in figure \ref{fig:Energy t} cannot be flat unless it has a different asymptotic behavior. The situation is different in case of Ellwood invariants in the tachyon approach (figure \ref{fig:Invariants tachyon}). With the exception of $E_1$ they are relatively stable only from  high levels, so we have very few points to do the fit and each of them makes a big difference. We estimate their statistical errors at high $t$ to be $0.1\sim 0.3$.


\subsection{Tachyon approach in OSFT}\label{sec:tachyon}

In the tachyon approach we have to start looking for nontrivial solutions at least at level 2, because at level 1 we have only the tachyon and marginal field and the only equation reads $\lS t=0$. At level 2 we find a pair of non-universal branches of solutions connected to the perturbative vacuum. Like in the toy model they have the same energy and opposite sign of $\lS$. These marginal branches end close to $t=0.9$, where they meet with a different pair of solutions (which also differs by a sign of $\lS$). The other branches have strongly negative energy (around $-350$ at $t=0$) and they are not a good starting point for level truncation.
The same story repeats at level 3. There the branches meet approximately at $t=1.8$ and the second branch has even more negative energy. Finally at level 4 the second branch gets a more reasonable energy and becomes a good seed for level truncation.

\begin{figure}[h]
\centering
\includegraphics[width=12cm]{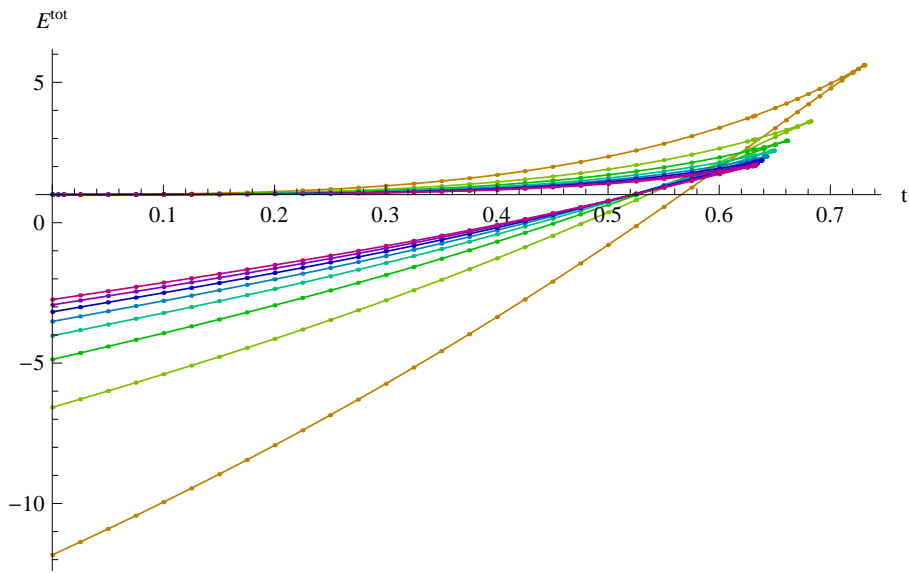}
\caption{Total energy of both branches of solutions in the tachyon approach.}\label{fig:En 2 branches}
\end{figure}

The total energy of both branches can be seen in figure \ref{fig:En 2 branches}. The branch structure is different from the toy model in section \ref{sec:toy tachyon}. In the toy model the marginal branches with opposite sign of $\lS$ ended when meeting each other. The situation here is more similar to the marginal approach \cite{large2}. This suggests that also the tachyon VEV could have a maximum as a function of $\l_B$, as it happens in the analytic example \cite{large-moduli} (the tachyon `bump' in fig. 3 there), but not in the toy model. However it is not possible to assess this, with the available data. We also reach the conclusion that the second branch is off-shell: for most of its length it has negative energy and closer inspection reveals that it does not satisfy neither the $t$-equation nor the out-of-Siegel equation\footnote{We compute just the first of the out-of-Siegel equations at level 2 and we denote it $\Delta_S$, see \cite{GaiottoRastelli}.}.

\begin{table}\nonumber
\centering
\begin{tabular}{|l|l|l|}\hline
L  & $t_{max}$ & $t_{\rm tv}$ \\\hline
2  & 0.892 & 0.5442 \\
4  & 0.730 & 0.5484 \\
6  & 0.682 & 0.5479 \\
8  & 0.661 & 0.5471 \\
10 & 0.649 & 0.5463 \\
12 & 0.642 & 0.5456 \\
14 & 0.638 & 0.5451 \\
16 & 0.634 & 0.5446 \\
18 & 0.632 & 0.5443 \\\hline
$\infty$ & 0.613 & 0.5405 \\\hline
\end{tabular}
\caption{The endpoint of the marginal branch in the tachyon approach with 3 digit precision. For comparison  we also show the tachyon coefficient of the tachyon vacuum solution at the same levels.}
\label{tab:endpoints}
\end{table}

In the rest of the section we concentrate only on the marginal branch. We show all data up to $t=0.632$, which is the endpoint of the branch at level 18. The general tendency is that the branch gets shorter by increasing the level\footnote{The decrease is not monotonous if we add odd levels.} (see table \ref{tab:endpoints}). Since the maximal value of the tachyon is bigger than the coefficient of the tachyon vacuum (which is also in table \ref{tab:endpoints}) it is indeed possible  that there is a bump in the tachyon.

\begin{figure}
\centering
\begin{subfigure}[t]{0.47\textwidth}
\includegraphics[width=\textwidth]{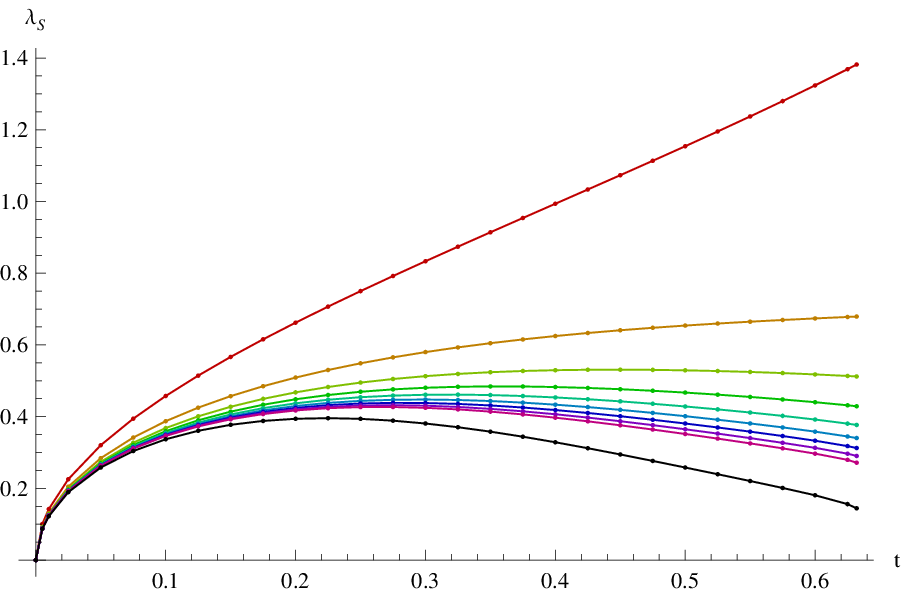}
\end{subfigure}\qquad
\begin{subfigure}[t]{0.47\textwidth}
\includegraphics[width=\textwidth]{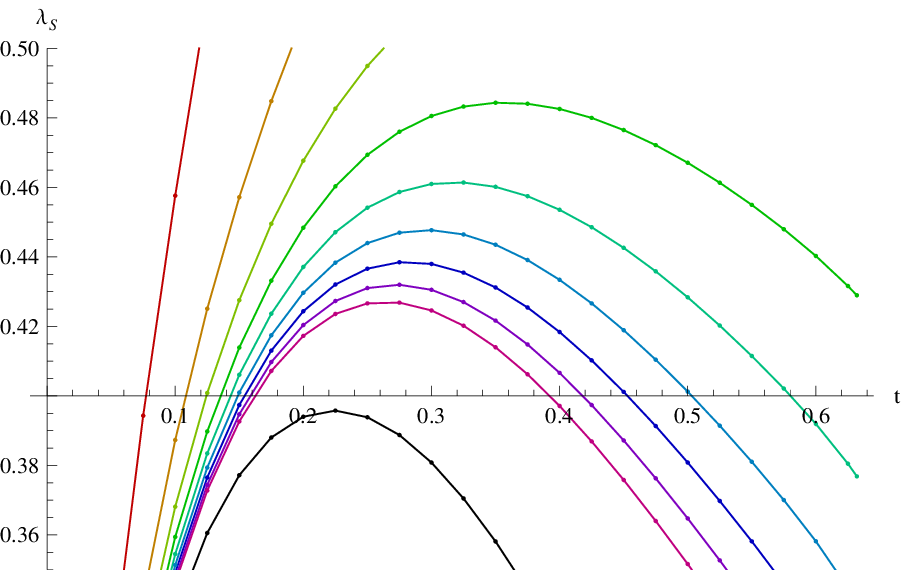}
\end{subfigure}
\caption{$\lS$ as a function of $t$ (left) and detail around the maximum of $\lS$ (right). The black line is infinite level fit.}\label{fig:lambda}
\end{figure}

When we inspect the relation between the marginal field and the tachyon (plotted in figure \ref{fig:lambda}) we find a maximum from level 5, so the marginal branch of the tachyon potential clearly covers a larger part of the moduli space than in the marginal approach. We will shortly see from the Ellwood invariants that we cover approximately twice as much. We show the critical value of the tachyon $t^\ast$ and the maximum of $\l_S$ in table \ref{tab:lambda star}. By increasing the level the critical $t^\ast$ gets smaller and the asymptotic value of the maximum $\l_S(t^\ast)$ is  $0.395\pm0.002$, which is  smaller than the length of the branch in the marginal approach, see \cite{large2} and below. This suggests that, as in the toy-model, part of the marginal branch in the marginal approach is off-shell. We will come back to this point later.

\begin{table}\nonumber
\centering
\begin{tabular}{|l|l|l|}\hline
L  & $t^\ast$ & $\lS^\ast$ \\\hline
6  & 0.444 & 0.531 \\
8  & 0.357 & 0.484 \\
10 & 0.319 & 0.461 \\
12 & 0.296 & 0.448 \\
14 & 0.282 & 0.439 \\
16 & 0.272 & 0.432 \\
18 & 0.264 & 0.427 \\\hline
$\infty$ & 0.223 & 0.395 \\\hline
$\sigma$ & 0.007 & 0.002 \\\hline
\end{tabular}
\caption{Table with maximal value of the marginal field $\lS^\ast$ and the corresponding tachyon $t^\ast$. The numbers are computed by finding maximum of polynomial interpolation of data in figure \ref{fig:lambda}. At levels 2 and 4 there is no maximum on the physical branch.}
\label{tab:lambda star}
\end{table}

It is not possible to establish with accuracy if the whole branch of the tachyon potential we found describes marginal deformations all the way to the end or whether it becomes off-shell at some point as in the toy model in the marginal approach. We do not observe any dramatic change in any quantity comparable to figure \ref{fig:toy lambda equation}, but everything gets slowly worse as we increase $t$. This situation may well change at higher level, but this is beyond our present reach.

\begin{figure}
\centering
\begin{subfigure}{0.47\textwidth}
\includegraphics[width=\textwidth]{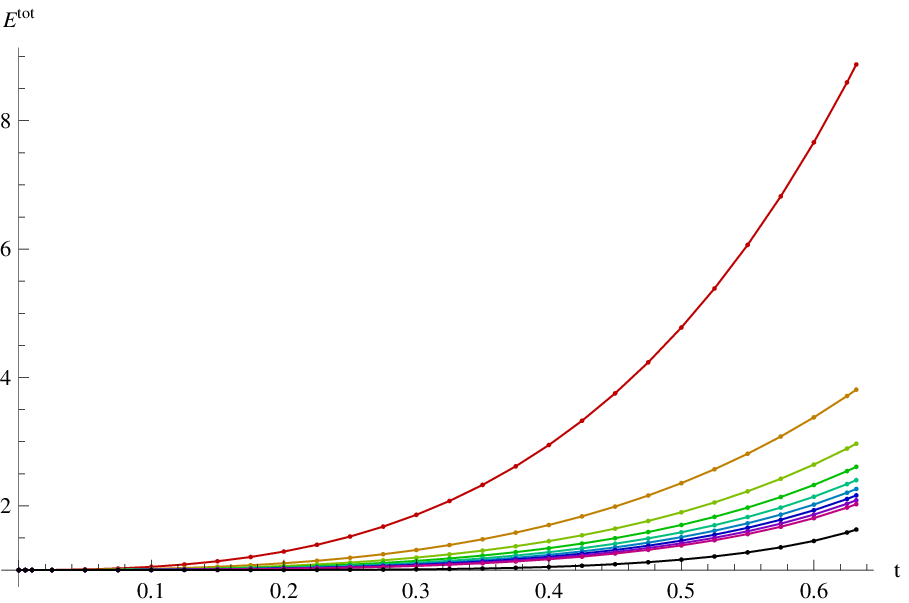}
\end{subfigure}\qquad
\begin{subfigure}{0.47\textwidth}
\includegraphics[width=\textwidth]{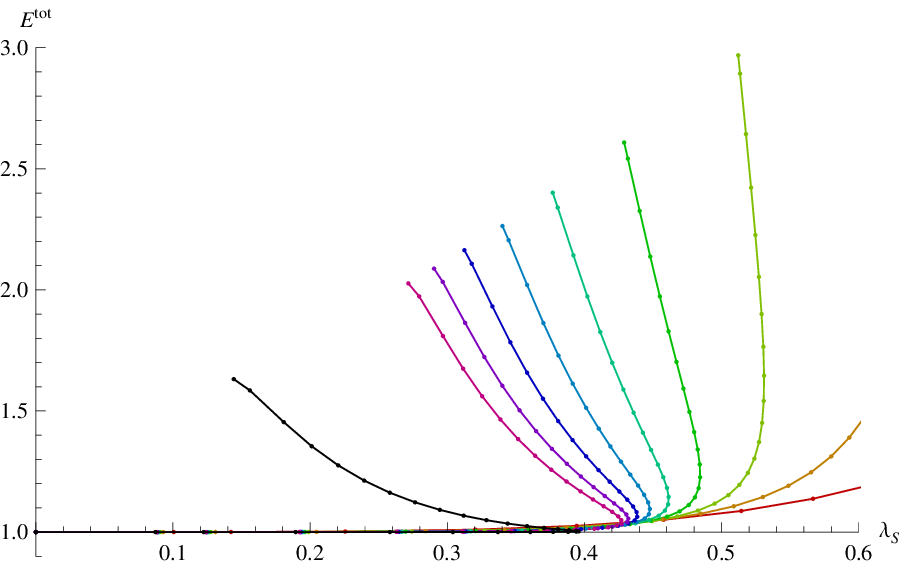}
\end{subfigure}
\begin{subfigure}{0.47\textwidth}
\includegraphics[width=\textwidth]{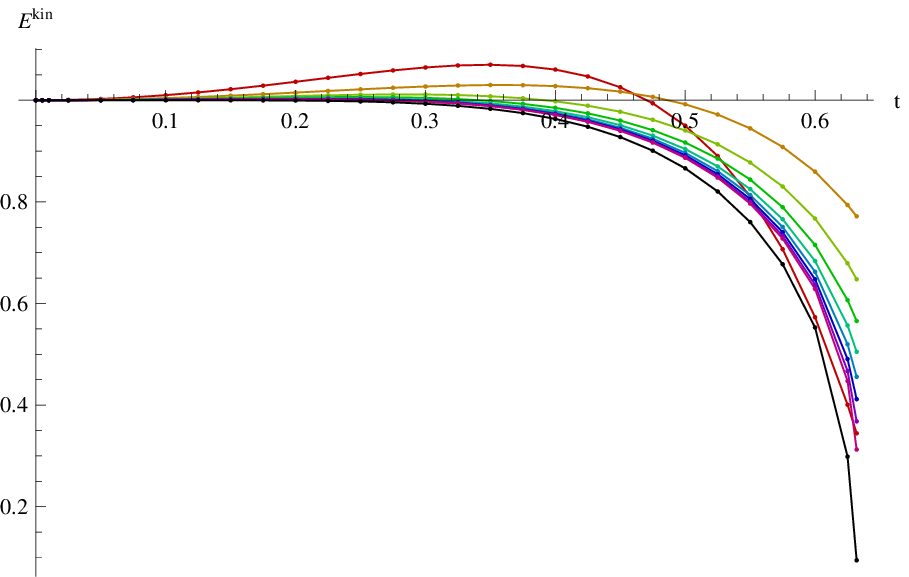}
\end{subfigure}\qquad
\begin{subfigure}{0.47\textwidth}
\includegraphics[width=\textwidth]{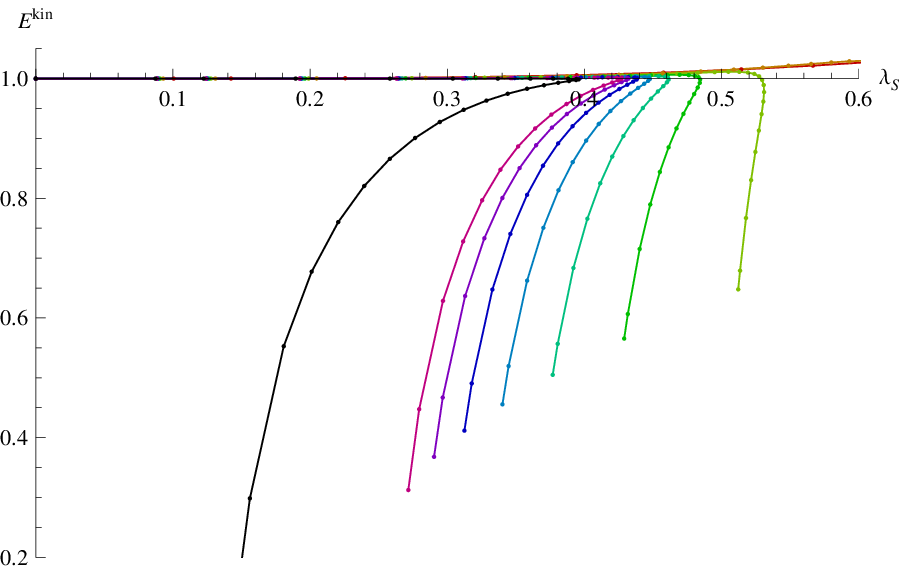}
\end{subfigure}
\begin{subfigure}{0.47\textwidth}
\includegraphics[width=\textwidth]{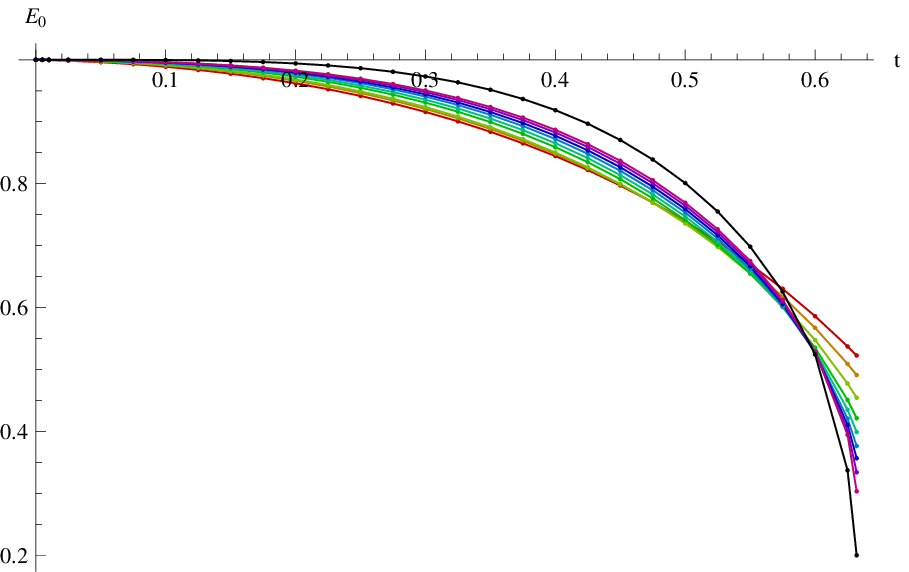}
\end{subfigure}\qquad
\begin{subfigure}{0.47\textwidth}
\includegraphics[width=\textwidth]{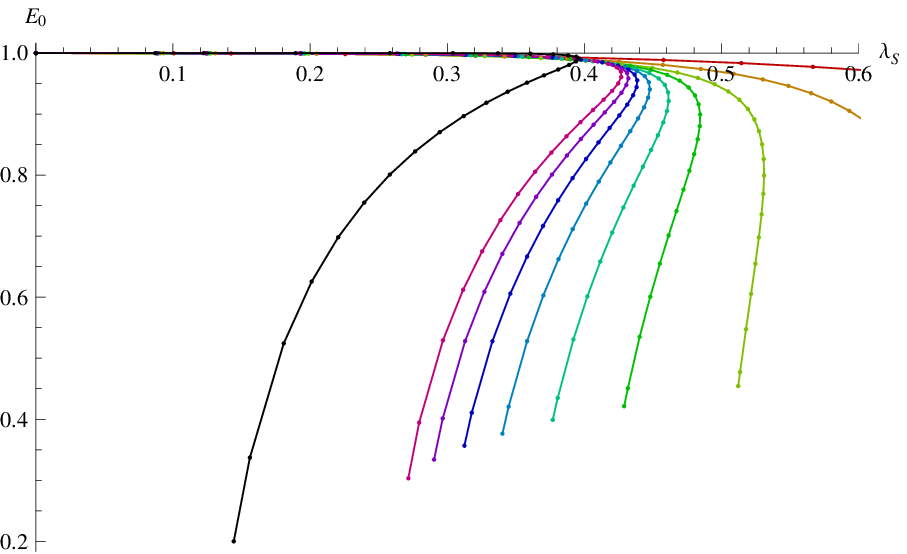}
\end{subfigure}
\caption{Energy of the marginal branch measured by the full action, by the kinetic term and by the $E_0$ invariant.
The figures on the right show energy as a function of $\lS$.}\label{fig:Energy t}
\end{figure}

We inspect the energy in the left part of figure \ref{fig:Energy t} and we compute both the full energy and energy from the kinetic term.\footnote{To be precise we define the total energy as $E^{tot}=1+2\pi^2\left(\frac{1}{2}\la\Psi,Q\Psi\ra+\frac{1}{3}\la\Psi,\Psi\ast\Psi\ra\right)$ and the kinetic energy as $E^{kin}=1+\frac{\pi^2}{3}\la\Psi,Q\Psi\ra$. Using this definition we find that the missing equation of motion is given by $\la 0|c_{-1} |Q\Psi+\Psi\ast\Psi\ra=\frac{3}{2\pi^2 t}(E^{tot}-E^{kin})$.} Their difference is proportional to $t$ times the missing equation of motion (see figure \ref{fig:missing eq tachyon}). Unlike the marginal approach, where the energy differs from 1 just by few percent (see section \ref{sec:marginal}), here it gets up to 2 at the end of the branch even at level 18 and the $t$-equation is badly violated there. The kinetic energy is monotonously decreasing and from $t\gtrapprox 0.3$ it is lower than 1. Therefore it seems unlikely that it can converge  to the correct value.
However comparing the values at different levels at the same $t$ does not seem to be the correct approach. We can instead compare with  the marginal approach and  look at the energy as a function of $\lS$ (see right part of figure \ref{fig:Energy t}), getting a different picture. We notice that at fixed $\lS$ both energies are close to 1 up to approximately $\lS^\ast$ and the kinetic energy is monotonously increasing in the second part of the branch. 

We can also measure the energy from the $E_0$ invariant \cite{Ishi} which is also plotted in figure \ref{fig:Energy t}. Its behavior is similar to $E^{kin}$, with the difference that it is always smaller than 1. Close to the end of the branch it gets worse by increasing the level and it also does not converge to the correct value at constant $t$. Same as for the energy, its behavior gets better when we look at it as a function of $\lS$.

\begin{figure}
\centering
\begin{subfigure}{0.47\textwidth}
\includegraphics[width=\textwidth]{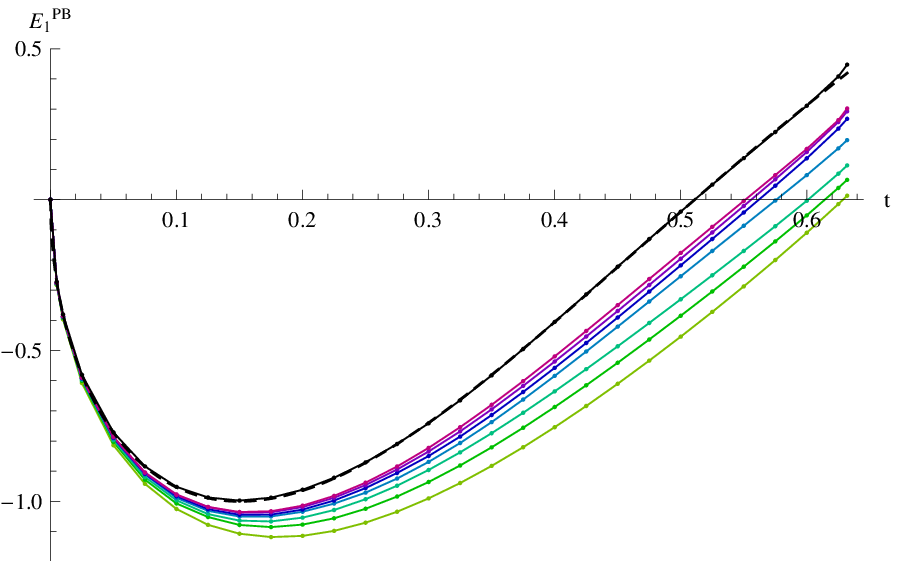}
\end{subfigure}\qquad
\begin{subfigure}{0.47\textwidth}
\includegraphics[width=\textwidth]{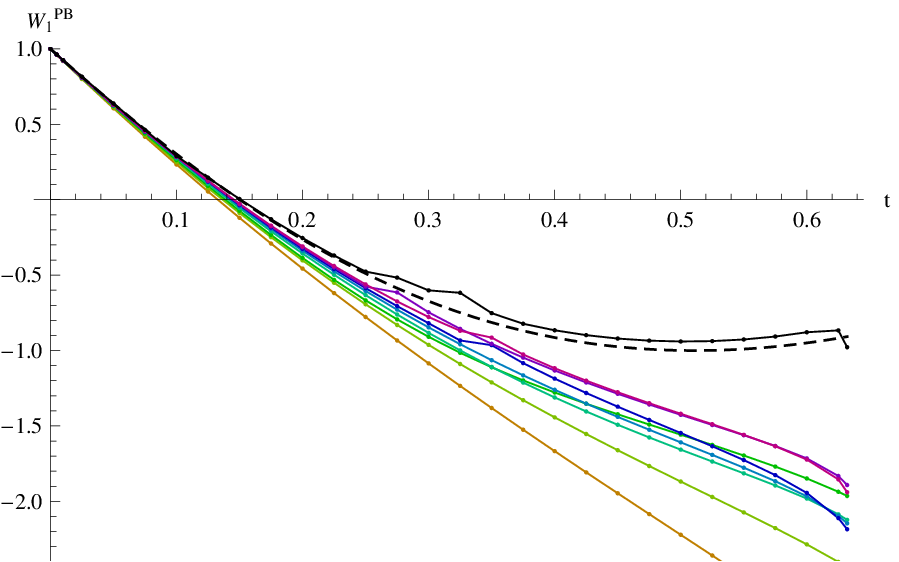}
\end{subfigure}
\begin{subfigure}{0.47\textwidth}
\includegraphics[width=\textwidth]{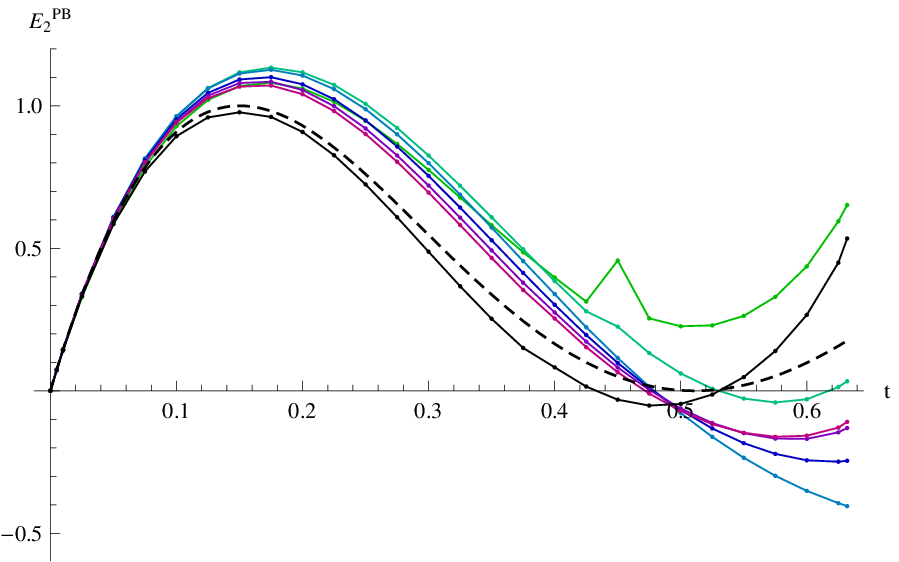}
\end{subfigure}\qquad
\begin{subfigure}{0.47\textwidth}
\includegraphics[width=\textwidth]{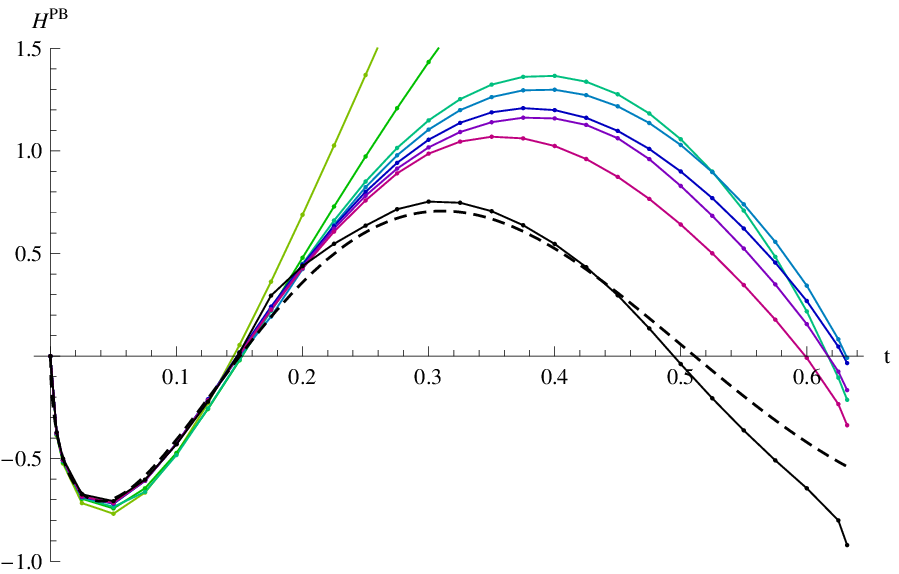}
\end{subfigure}
\begin{subfigure}{0.47\textwidth}
\includegraphics[width=\textwidth]{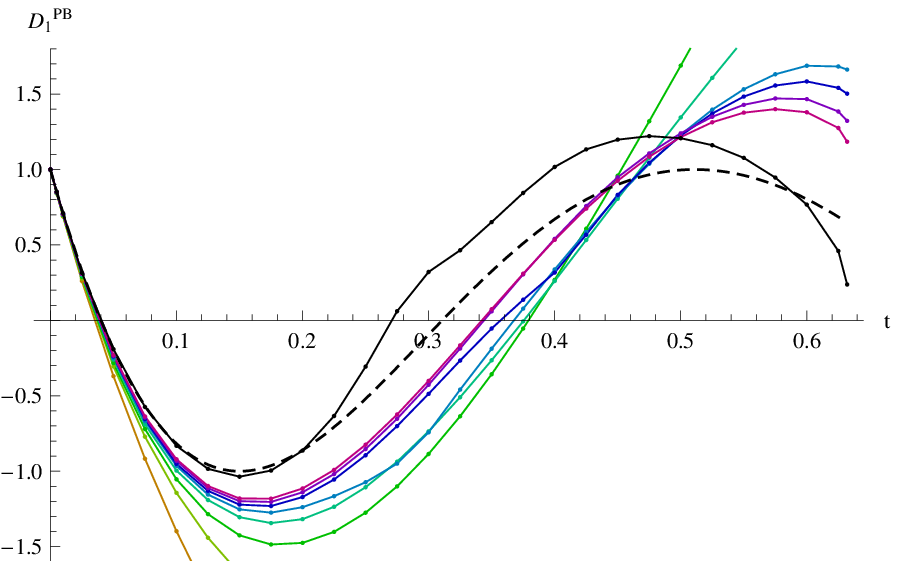}
\end{subfigure}\qquad
\begin{subfigure}{0.47\textwidth}
\includegraphics[width=\textwidth]{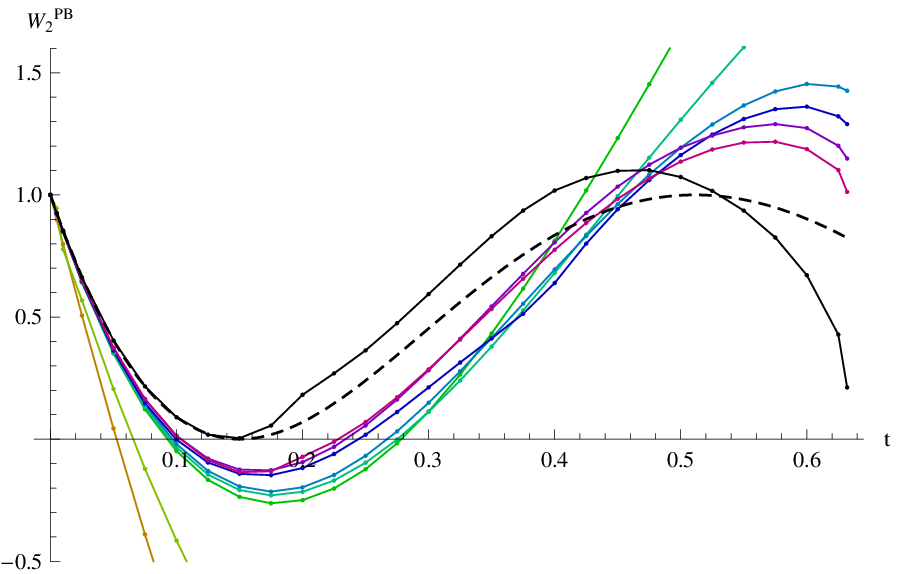}
\end{subfigure}
\caption{The six nontrivial invariants in the tachyon approach. We use Pad\'{e}-Borel approximation to improve the convergence. The black line is a linear fit to level $\infty$, the dashed line is an expected value based on order 4 $\lB$ fit (\ref{lambdaBCFT tachyon}). We removed some of the low level data which behaved too chaotically under the Pad\'{e}-Borel approximation.}
\label{fig:Invariants tachyon}
\end{figure}

Next we concentrate on the rest of the Ellwood invariants. There are six invariants that can tell us more about the relation between $t$ and $\lB$. The first momentum invariant $E_1$ is the best behaved and it is  enough to extract the BCFT modulus $\lB$. The remaining ones behave worse than in the marginal approach and the weight 1 invariants ($D_1$, $E_2$, $W_2$, $H$) oscillate quite a lot for large $t$ (the amplitude is bigger than the expected values by one order). Therefore we use Pad\'{e}-Borel resummation to improve their convergence. After that we get stable behavior from level 12. The infinite level extrapolation of the invariants stays reasonably well within the allowed range.

To fit the relation between $t$ and $\lB$ we recall that we must have $t\sim \lS^2+O(\lS^4)=\lB^2+O(\lB^4)$ from the perturbative construction of the solution. This suggests that we should use an expansion
\begin{equation}\label{lambdaBCFT tachyon}
\lB(t)=\sqrt{t}\ \sum_{i=0}^M a_i t^i.
\end{equation}
This polynomial ansatz cannot capture the behavior around the possible tachyon maximum, but since the solution is either off-shell or very imprecise at the end of the branch, we cannot describe this region well anyway.

\begin{figure}
\centering
\includegraphics[width=8cm]{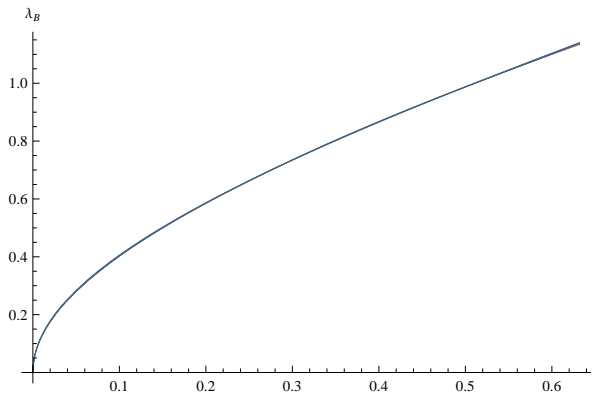}
\caption{$\lB$ as a function of  the tachyon.  The appearing curve is in fact the superposition of five different orders fits ($M=1,2,3,4,5$).  This shows that the $\lB$ vs $t$ relation  (\ref{lambdaBCFT tachyon}) is
 essentially insensitive the fit's order $M$.}\label{fig:lambdaB_tach}
\end{figure}

\begin{figure}
\centering
\begin{subfigure}[t]{0.47\textwidth}
\includegraphics[width=\textwidth]{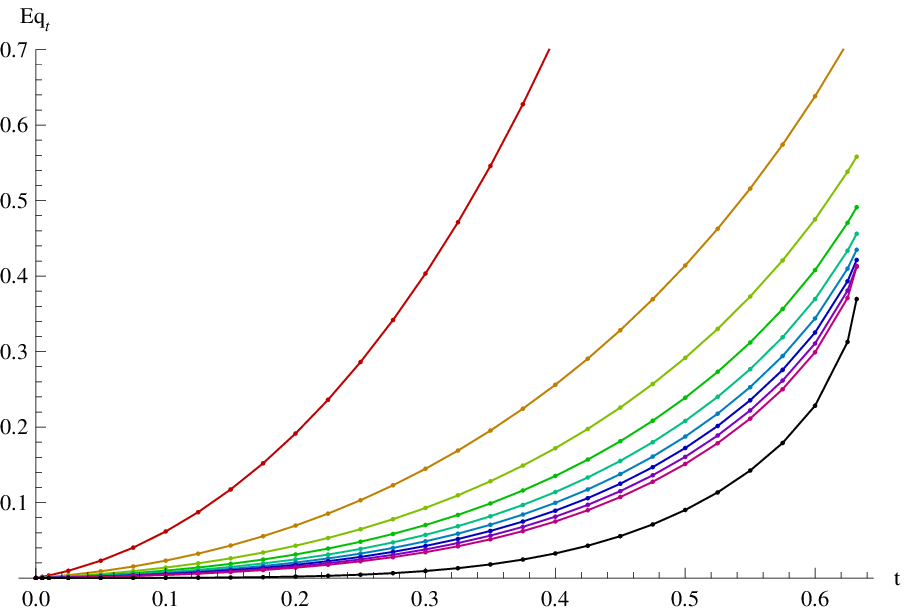}
\end{subfigure}\qquad
\begin{subfigure}[t]{0.47\textwidth}
\includegraphics[width=\textwidth]{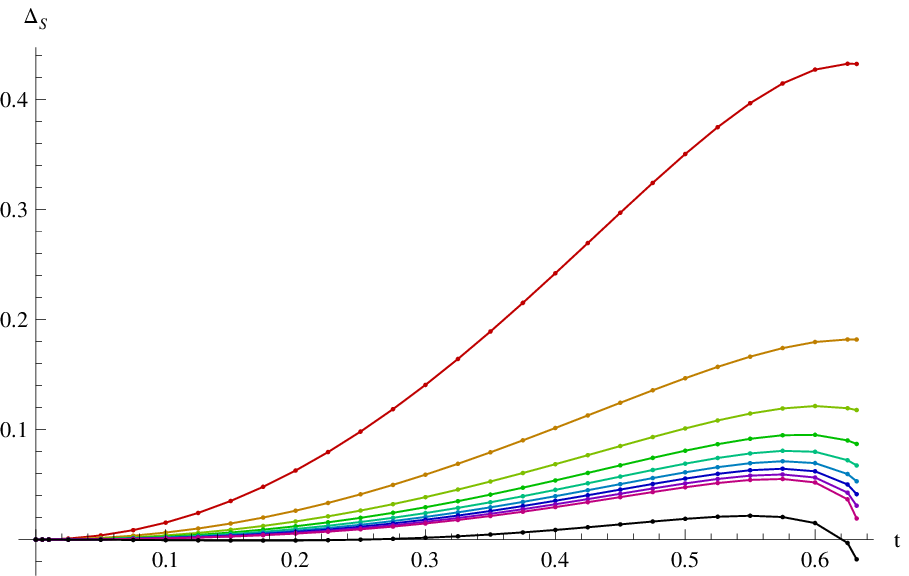}
\end{subfigure}
\caption{Missing equation for $t$ (left) and first out-of-Siegel equation $\Delta_S$ (right) in the tachyon approach.}\label{fig:missing eq tachyon}
\end{figure}

We have tried several different methods of fitting $\lB$. The best we found  uses just  the $E_1$ invariant. Only this invariant is stable from level 6 and it is free from  anomalous behavior in the Pad\'{e}-Borel approximation. We determine the numbers $a_i$ by minimizing the sum of  differences between the infinite level fit and the expected behavior based on (\ref{lambdaBCFT tachyon}) and (\ref{Elw expected}), with weights given by the missing equation (in order to give more importance to the points where the full OSFT equation of motion is better satisfied). Although the $a_i$ with $i\geq2$ are not very stable with respect to the order $M$ and other parameters of the fit, the $\lB$ functions are very similar in the allowed range of $t$. We find that $a_0\approx1.23$ and $a_1\approx0.4$, see table \ref{tab:fit lambdaB t}, fig. \ref{fig:lambdaB_tach}.

\begin{table}\nonumber
\centering
\begin{tabular}{|l|llllll|}\hline
$M$ & $a_0$   & $a_1$    & $a_2$     & $a_3$    & $a_4$    & $a_5$ \\\hline
1   & 1.25521 & 0.283323 &           &          &          & \\
2   & 1.23615 & 0.395288 & -0.147794 &          &          & \\
3   & 1.23365 & 0.434811 & -0.286707 & 0.137257 &          & \\
4   & 1.23777 & 0.306322 &  0.499518 & -1.58704 & 1.25907  & \\
5   & 1.23395 & 0.470130 & -1.00362  & 3.91381  & -7.58098 & 5.20085 \\\hline
\end{tabular}
\caption{Parameters of the $\lB$ fit (\ref{lambdaBCFT tachyon}) of order  $M$ up to 5.}
\label{tab:fit lambdaB t}
\end{table}


We show the invariants in figure \ref{fig:Invariants tachyon}. The $E_1$ invariant almost perfectly matches the $\lB$ fit. The $W_1$ invariant also matches very well. The  $E_2$ and $H$ invariants agree the fit quite well up to high $t$ as well. The rest of the invariants ($D_1$ and $W_2$) follow the fit well at small $t$, but at some point they suddenly move away from it.   This effect is caused by  a combination of the Pad\'{e}-Borel approximation and the infinite level fit (see the not very smooth curves of figure \ref{fig:Invariants tachyon}), but since we have very few data points we are unable to eliminate it.

We can see that the invariants of the solution cover more than two fundamental domains, so using the individual invariants we can find the values of tachyon and marginal field that correspond to Dirichlet boundary condition ($\lB=1/2$) and Neumann boundary condition with Wilson line ($\lB=1$),  which is the point in moduli space that is most distant to the perturbative vacuum. The results are in table \ref{tab:lambda D N}. For the Dirichlet boundary condition we find with a good precision $t^D=0.152$ and $\lS^D=0.378$. The results for Neumann boundary conditions have  larger errors and depend much more on the chosen invariant, but  the numbers from $E_1$ are the most reliable, since they are the least contaminated by numerical effects. Notice that the computed values (with the corresponding error estimated by varying the fit parameters) for the Neumann point  are not entirely consistent between  themselves. This is because there are  extra errors induced by the Pad\'e-Borel approximation which we cannot estimate and which mostly affect the other invariants $W_1, D_1, E_2, W_2, H$ at high $t$. 
\begin{table}\nonumber
\centering
\begin{tabular}{|l|ll|ll|}\hline
Invariant & $t^D$           & $\lS^D$         & $t^N$           & $\lS^N$         \\\hline
$E_1$     & 0.152$\pm$0.002 & 0.378$\pm$0.003 & 0.504$\pm$0.005 & 0.255$\pm$0.005 \\\hline
$W_1$     & 0.152$\pm$0.001 & 0.378$\pm$0.002 & 0.46 $\pm$0.03  & 0.29 $\pm$0.02  \\
$D_1$     & 0.150$\pm$0.002 & 0.377$\pm$0.002 & 0.46$ \pm$0.01  & 0.29 $\pm$0.01 \\
$E_2$     & 0.151$\pm$0.001 & 0.378$\pm$0.003 & 0.481$\pm$0.005 & 0.272$\pm$0.005 \\
$W_2$     & 0.145$\pm$0.005 & 0.375$\pm$0.004 & 0.45$ \pm$0.01  & 0.29 $\pm$0.01  \\
$H  $     & 0.150$\pm$0.004 & 0.377$\pm$0.003 & 0.49$ \pm$0.03  & 0.27 $\pm$0.02 \\\hline
\end{tabular}
\caption{Table with $t$ and $\lS$ that correspond to Dirichlet and Neumann boundary condition. The numbers are found from interpolation of infinite level fit. }
\label{tab:lambda D N}
\end{table}

\FloatBarrier
\subsection{Marginal approach in OSFT} \label{sec:marginal}
\begin{figure}[t]
\centering
\begin{subfigure}{0.47\textwidth}
\includegraphics[width=\textwidth]{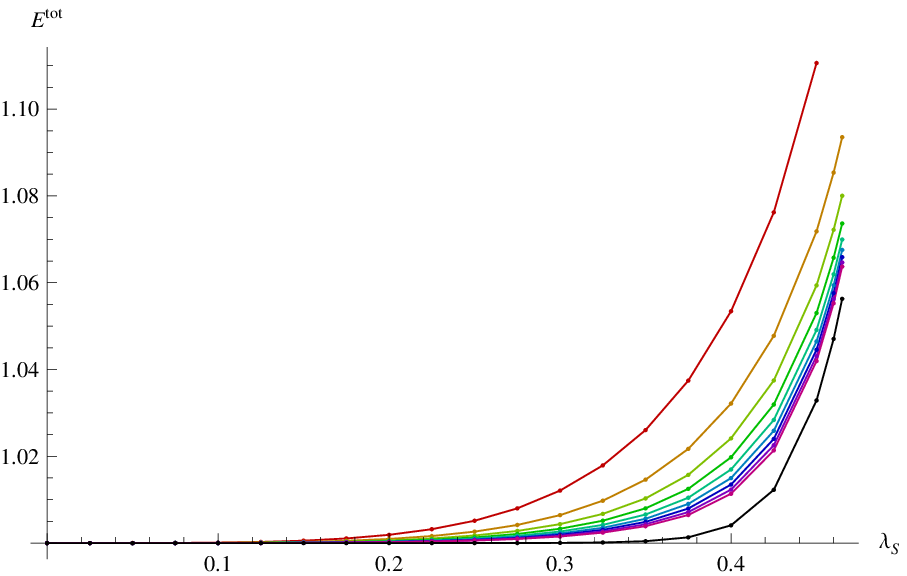}
\end{subfigure}\qquad
\begin{subfigure}{0.47\textwidth}
\includegraphics[width=\textwidth]{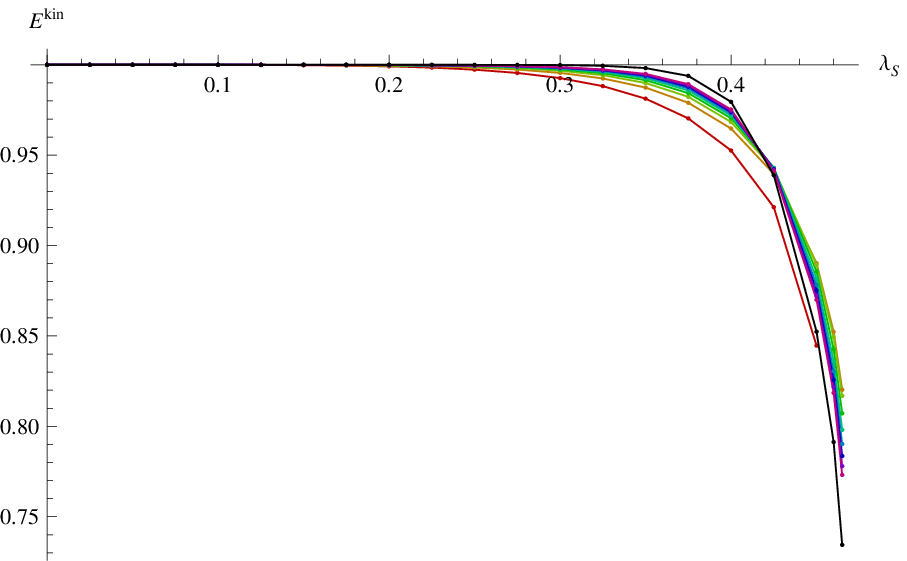}
\end{subfigure}
\begin{subfigure}{0.47\textwidth}
\includegraphics[width=\textwidth]{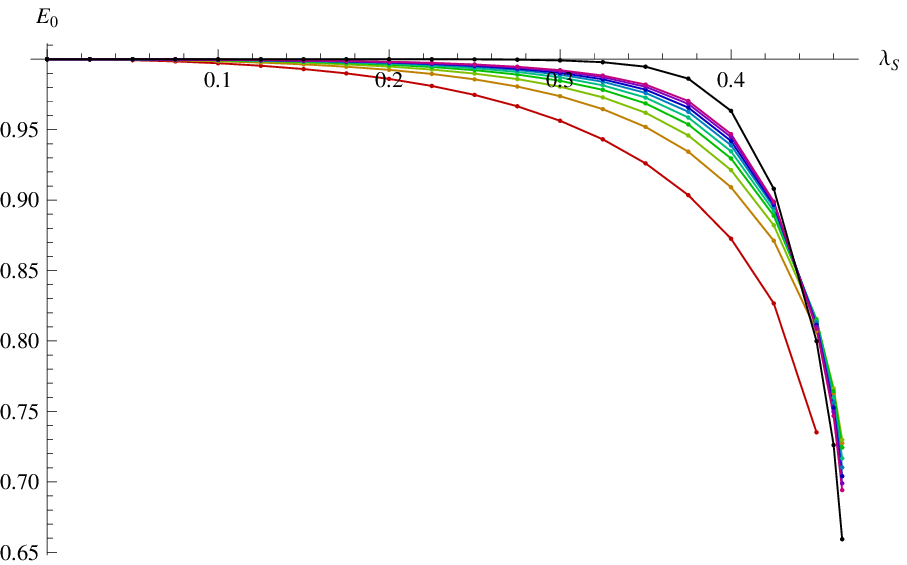}
\end{subfigure}
\caption{Energy measured by $E^{tot}$, $E^{kin}$ and $E_0$ in the marginal approach.}\label{fig:Energy m}
\end{figure}

\begin{figure}[t]
\centering
\begin{subfigure}{0.47\textwidth}
\includegraphics[width=\textwidth]{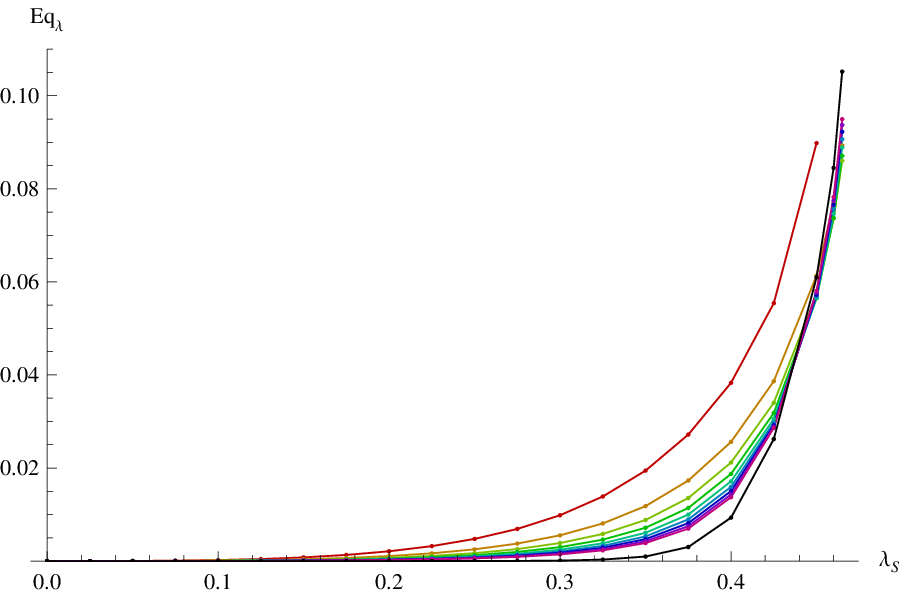}
\end{subfigure}\qquad
\begin{subfigure}{0.47\textwidth}
\includegraphics[width=\textwidth]{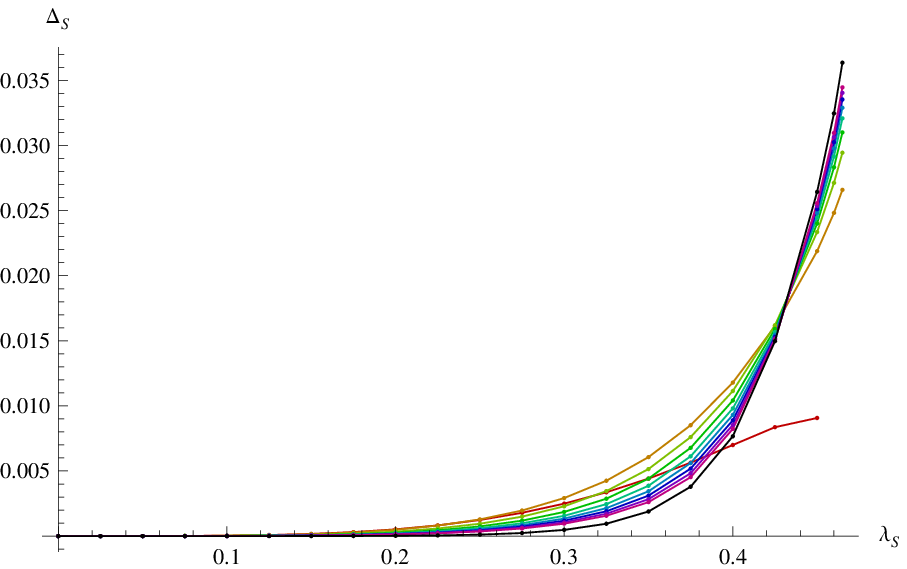}
\end{subfigure}
\caption{Missing equation for $\lS$ (left) and first out-of-Siegel equation $\Delta_S$ (right) in the marginal approach.}\label{fig:missing eq marginal}
\end{figure}

\begin{figure}
\centering
\begin{subfigure}{0.47\textwidth}
\includegraphics[width=\textwidth]{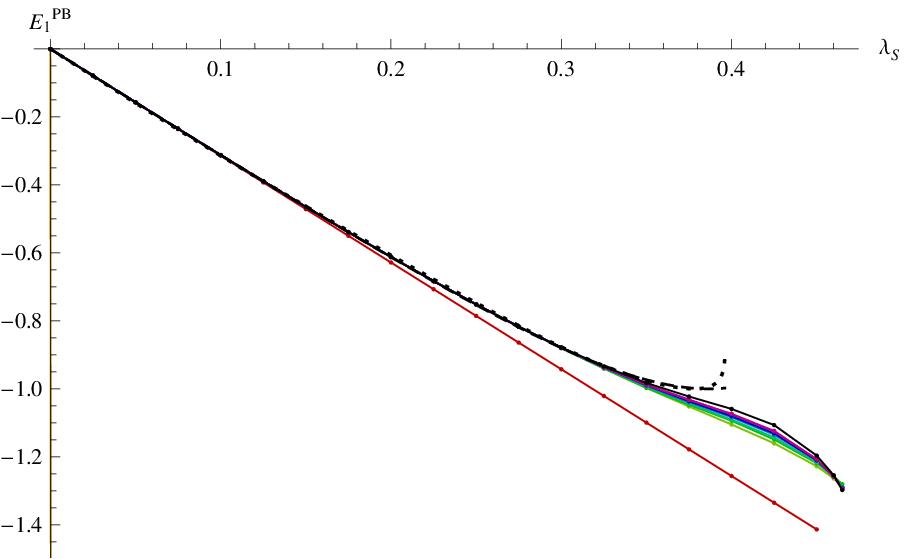}
\end{subfigure}\qquad
\begin{subfigure}{0.47\textwidth}
\includegraphics[width=\textwidth]{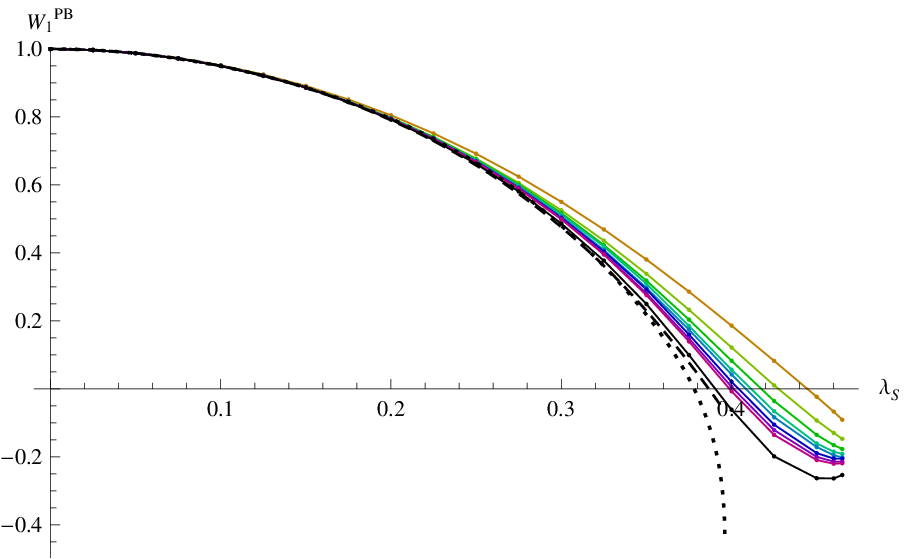}
\end{subfigure}
\begin{subfigure}{0.47\textwidth}
\includegraphics[width=\textwidth]{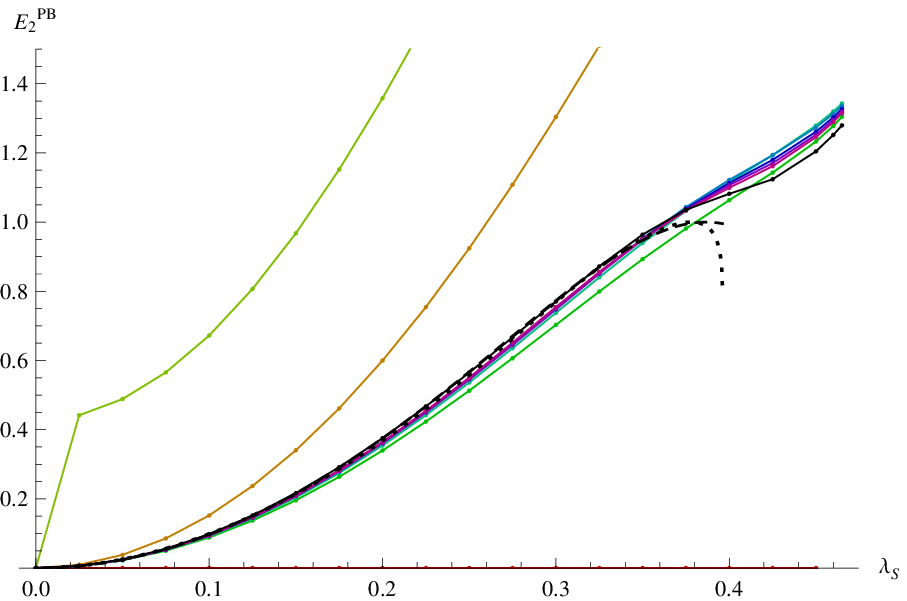}
\end{subfigure}\qquad
\begin{subfigure}{0.47\textwidth}
\includegraphics[width=\textwidth]{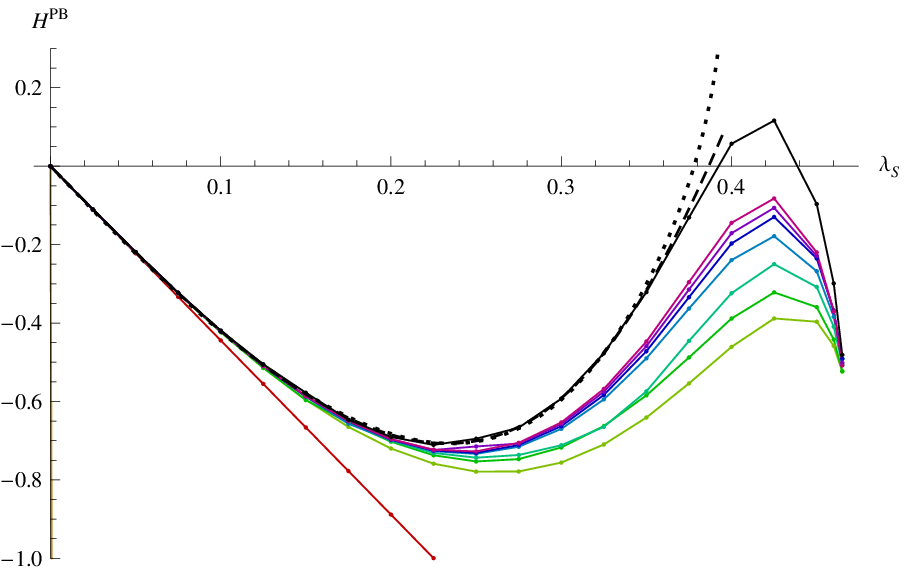}
\end{subfigure}
\begin{subfigure}{0.47\textwidth}
\includegraphics[width=\textwidth]{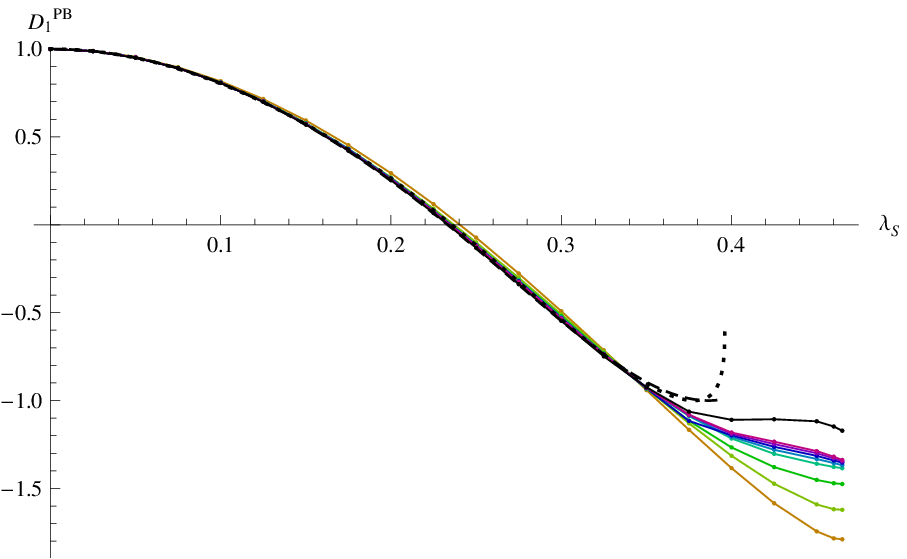}
\end{subfigure}\qquad
\begin{subfigure}{0.47\textwidth}
\includegraphics[width=\textwidth]{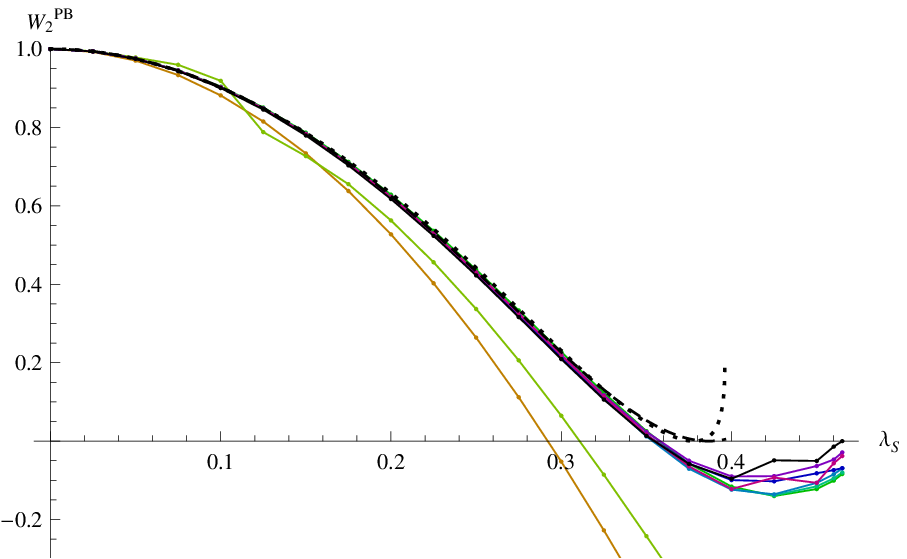}
\end{subfigure}
\caption{Pad\'{e}-Borel approximation of six nontrivial invariants in marginal approach. The black line is a linear fit to level $\infty$.
The dashed line is an expected value based on $\lB$ fit of order 3 up to $\lS=0.3$, (\ref{lambdaBCFT marginal}). The dotted line is an expected value based on $\lB$ fit from the tachyon approach, (\ref{lambdaBCFT marginal2}).}\label{fig:Invariants marginal}
\end{figure}

In this section we provide improved data for the marginal solution from \cite{large2}. We increase the level by 6 and we add the whole set of converging Ellwood invariants. The new levels give only a slight improvement in energy and the missing equations (figure \ref{fig:Energy m}  and \ref{fig:missing eq marginal}) but the new invariants give us a more complete understanding of the properties of the solution, see figure \ref{fig:Invariants marginal}.


First we determine the approximate relation between $\lS$ and $\lB$ as in the tachyon approach. This time we can use a polynomial ansatz
\begin{equation}\label{lambdaBCFT marginal}
\lB(\lS)=\sum_{i=1}^M a_i \lS^{2i-1},
\end{equation}
notice that there are only odd powers because the function must be odd in $\l_S$. We fit the parameters by minimizing the sum of differences between the extrapolated invariants and the expected values based on (\ref{Elw expected}) and (\ref{lambdaBCFT marginal}). We know that $a_1=1$ from the perturbative expansion of the solution. This expansion works well only for small $\lS$ and it fails around the maximum of $\lS$, where the perturbative series stop converging. 
Like in the tachyon approach some of the coefficients $a_i$ vary with $M$, but the final functions are not very different. The $a_1$ coefficient is always very close to 1 as we expected and the $a_2$ coefficient is also approximately 1. The results using data up to $\lS=0.3$ with different $M$ are in table \ref{tab:fit lambdaB m}.

\begin{table}[t]\nonumber
\centering
\begin{tabular}{|l|lllll|}\hline
$M$ & $a_1$   & $a_2$    & $a_3$   & $a_4$    & $a_5$ \\\hline
2   & 0.98404 & 1.68599  &         &          & \\
3   & 1.00075 & 0.931354 & 6.8701  &          & \\
4   & 0.99836 & 1.13846  & 2.26019 & 29.3492  & \\
5   & 1.00035 & 0.854005 & 13.6089 & -139.462 & 836.082 \\\hline
\end{tabular}
\caption{Parameters of the $\lB$ fit (\ref{lambdaBCFT marginal}) of several different orders.}
\label{tab:fit lambdaB m}
\end{table}

To compare  and to get an handle on the region close to maximum of $\lS$ we have fitted the data from the tachyon approach with the following function
\begin{equation}\label{lambdaBCFT marginal2}
\lB(\lS)=\frac{2(\lS^\ast)^2}{\lS^2}\left(1-\sqrt{1-\frac{\lS^2}{(\lS^\ast)^2}} \right)\sum_{i=1}^M a_i \lS^{2i-1},
\end{equation}
 and we add it to figure \ref{fig:Invariants marginal} for comparison.

The numerical results and the fit agree quite well at least up to $\lS\sim 0.35$ and then they deviate. Recall that in the tachyon approach we found that the maximum of $\l_S$ is $\sim 0.4$. Consistently, for $\lS>0.4$ the invariants do not behave according to their $\sin/\cos$ dependency, so in this region the solution is clearly off-shell, as it happens in the toy-model. When we compare the results to the prediction from the tachyon approach we clearly miss the part after the Dirichlet point where the invariants would have to go vertically. It is in fact possible that the solution becomes off-shell at the Dirichlet point or slightly before it (therefore at a smaller value than the $\lS^\ast$ computed from the tachyon approach), but the observed behavior does not allow us to determine the position of the Dirichlet point with the same good  precision as in the tachyon approach.  


\section{Conclusions}
In this paper we have tested level truncation and Siegel gauge against large marginal boundary deformations of an initial BCFT$_0$. We have ``experimentally'' observed that, in Siegel gauge as well, the coefficient of the marginal field $\l_S$ in the level-truncated OSFT solution is not injectively related to the BCFT modulus $\l_B$ but it has a maximum, compatible with the critical value where the solution stop existing in the standard marginal approach. To achieve this we have  used the tachyon coefficient of the marginal solution as a coordinate in the BCFT moduli space and we have found that a much larger region of the BCFT moduli space is covered in this way. In particular the Ellwood invariants computed from the level-truncated solution cover the full periodic  moduli space of the cosine deformation. However the missing equations of motion get worse as we increase the VEV of the tachyon and the quite high level we reached is still not enough to flatten the new-found branch of the tachyon potential, after $\lS$ reaches its maximum.

Taking inspiration from the toy model, we have also made considerable effort in searching for other marginal branches that would describe $\lB>1/2$ in the standard marginal approach, but so far without success. We took solutions from the tachyon approach with $t>t^\ast$ and used them as starting points in the marginal approach at the same level and $\lS$ but we fell in the basin of attraction of the known solutions.
We have also scanned all solutions up to level 4, where we have 20 fields and around half a million of solutions, as possible starting points. We have checked the solutions for several different values of $\lS$, however there is no new solution that would describe marginal deformations.
Finding all solutions at higher level is not concretely viable with the available hardware and software facilities, which is quite unfortunate given that our tachyon approach suggests that the new branch may appear at level 5 (36 fields, about 34 billions of solutions) where, for the first time, $\lS$ has a maximum.

The consistent behaviour of the first Ellwood invariants in the tachyon approach (in particular the first momentum invariant  $E_1$) are  positive indications in favour of a full Siegel-gauge solution reaching and going past $\lB=1$ (corresponding to the Neumann point with Wilson line of the cosine deformation), but other data (particularly  the behaviour of the missing equation of motion in the tachyon approach and the yet not-observed new branch in the marginal approach) suggest more caution. To be conservative, we cannot exclude that the exact Siegel gauge solution may just stop at the found maximum of $\lS$.  In this case a deformation of the gauge condition would be needed to displace the solution further in moduli space, and this possibility should  be also  considered.

Now,  more effort should  be devoted to the (vastly unknown) analytic description of this  important corner of the OSFT landscape.
\section*{Acknowledgments}

We thank Ted Erler, Tom\'a\v{s} Proch\'azka, Martin Schnabl and Roberto Tateo for useful discussions.
CM thanks the Academy of Science of Czech Republic for kind hospitality and support during the beginning  of this work and the organizers of the  workshop ``Gauge theories, supergravity and superstrings'' in Benasque, for providing a stimulating environment during part of this research.

 The research of MK has been supported by Grant Agency of the Czech Republic, under the grant 14-31689S.
The research of CM is funded by a {\it Rita Levi Montalcini} grant from the Italian MIUR. 
Computational resources were provided by the MetaCentrum under the program LM2010005 and the CERIT-SC under the program Centre CERIT Scientific Cloud,
part of the Operational Program Research and Development for Innovations, Reg. no. CZ.1.05/3.2.00/08.0144.

\appendix

\section{Some periodic solutions to the Lam\'e equation} \label{app:Lame}
Here we write the polynomial solutions to the $j=3$ Lam\'e equation, written in the form
\begin{equation}
-\frac{d^2\psi_n}{dy^2}+12 m\ \sn^2 (y|m)\psi_n(y)=\left(4+4m+4\Delta_m M^2_n\right)\psi_n(y).
\end{equation}
There are $7=2j+1$ polynomial solutions and all the polynomial solutions must be of order $j$ in the elliptic functions \cite{BatemanIII}.

\begin{table}[h]\nonumber
\centering
\begin{tabular}{|l|l|l|l|l|}\hline
$n$ & Solution              & $4\Delta_m M^2_n$       & $a_n$                                             & Period  \\\hline
1   & $\sn y\ \cn y\ \dn y$ & $0                   $  & -                                                 & $2K(m)$ \\
2   & $\dn^3 y+a_2\ \dn y$  & $-2+m-2\sqrt{1-m+4m^2}$ & $\frac{1}{5}\left(-4+2m+\sqrt{1-m+4m^2}\right)$   & $2K(m)$ \\
3   & $\dn^3 y+a_3\ \dn y$  & $-2+m+2\sqrt{1-m+4m^2}$ & $\frac{1}{5}\left(-4+2m-\sqrt{1-m+4m^2}\right)$   & $2K(m)$ \\\hline
4   & $\sn^3 y+a_4\ \sn y$  & $1+m-2\sqrt{4-7m+4m^2}$ & $-\frac{1}{5m}\left(2+2m+\sqrt{4-7m+4m^2}\right)$ & $4K(m)$ \\
5   & $\sn^3 y+a_5\ \sn y$  & $1+m+2\sqrt{4-7m+4m^2}$ & $-\frac{1}{5m}\left(2+2m-\sqrt{4-7m+4m^2}\right)$ & $4K(m)$ \\
6   & $\cn^3 y+a_6\ \cn y$  & $1-2m-2\sqrt{4-m+m^2}$  & $\frac{1}{5m}\left(2-4m+\sqrt{4-m+m^2}\right)$    & $4K(m)$ \\
7   & $\cn^3 y+a_7\ \cn y$  & $1-2m+2\sqrt{4-m+m^2}$  & $\frac{1}{5m}\left(2-4m-\sqrt{4-m+m^2}\right)$    & $4K(m)$ \\\hline
\end{tabular}
\caption{Polynomial solutions to the  Lam\'e equation. The first three have the correct periodicity for our Schr\"{o}dinger problem in section \ref{sec:toy} and they are the bound states of the Schr\"{o}dinger potential.}
\label{tab:Lame}
\end{table}

\end{document}